\documentclass[aps,prb,twocolumn,superscriptaddress,showpacs]{revtex4-1}
\usepackage{amssymb, textcomp}
\usepackage{graphicx}
\usepackage{hyperref}
\usepackage{siunitx}

\newcommand{\degree}{\ensuremath{^\circ}}

\begin{document}
\title{Domain wall induced spin-polarized flat bands in antiferromagnetic topological insulators}

\author{E. K. Petrov}
\email[]{evg.konst.petrov@gmail.com}
\affiliation{Tomsk State University, 634050 Tomsk, Russia}
\affiliation{St. Petersburg State University, 199034 St. Petersburg,  Russia}

\author{V. N. Men'shov}
\affiliation{NRC Kurchatov Institute, 123182 Moscow, Russia}
\affiliation{Tomsk State University, 634050 Tomsk, Russia}
\affiliation{St. Petersburg State University, 199034 St. Petersburg,  Russia}

\author{I. P. Rusinov}
\affiliation{Tomsk State University, 634050 Tomsk, Russia}
\affiliation{St. Petersburg State University, 199034 St. Petersburg,  Russia}

\author{M. Hoffmann}
\affiliation{Institute for Theoretical Physics, Johannes Kepler University, 4040 Linz, Austria}

\author{A. Ernst}
\affiliation{Institute for Theoretical Physics, Johannes Kepler University, 4040 Linz, Austria}
\affiliation{Max Planck Institute of Microstructure Physics, 06120 Halle, Germany}

\author{M. M. Otrokov}
\affiliation{Centro de F\'{\i}sica de Materiales (CFM-MPC), Centro Mixto CSIC-UPV/EHU, 20018 Donostia-San Sebasti\'{a}n, Basque Country, Spain}
\affiliation{IKERBASQUE, Basque Foundation for Science, 48011 Bilbao, Spain}

\author{V. K. Dugaev}
\affiliation{Department of Physics and Medical Engineering, Rzesz\'ow University of Technology, 35-959 Rzesz\'ow, Poland}

\author{T. V. Menshchikova}
\affiliation{Tomsk State University, 634050 Tomsk, Russia}

\author{E. V. Chulkov}
\email[]{evguenivladimirovich.tchoulkov@ehu.eus}
\affiliation{Departamento de F\'{\i}sica de Materiales, Facultad de Ciencias Qu\'{\i}micas, Universidad del Pa\'{\i}s Vasco, Apdo. 1072, 20080 San Sebasti\'{a}n/Donostia, Spain}
\affiliation{Donostia International Physics Center (DIPC), 20018 San Sebasti\'{a}n/Donostia, Spain}
\affiliation{St. Petersburg State University, 199034 St. Petersburg,  Russia}
\affiliation{Centro de F\'{\i}sica de Materiales (CFM-MPC), Centro Mixto CSIC-UPV/EHU, 20018 Donostia-San Sebasti\'{a}n, Basque Country, Spain}
\affiliation{Tomsk State University, 634050 Tomsk, Russia}

\newcommand*{\vv}[1]{\textbf{#1}}
\newcommand*{\BT}{Bi$_2$Te$_3$}
\newcommand*{\BS}{Bi$_2$Se$_3$}
\newcommand*{\ST}{Sb$_2$Te$_3$}
\newcommand*{\BTS}{Bi$_2$Te$_2$Se}
\newcommand*{\MBS}{MnBi$_2$Se$_4$}
\newcommand*{\MBT}{MnBi$_2$Te$_4$}
\newcommand*{\MBTS}{MnBi$_2$Te$_2$Se$_2$}
\newcommand*{\VBS}{VBi$_2$Se$_4$}
\newcommand*{\VBTS}{VBi$_2$Te$_2$Se$_2$}
\newcommand*{\VBT}{VBi$_2$Te$_4$}
\newcommand*{\VST}{VSb$_2$Te$_4$}
\newcommand*{\VSTS}{VSb$_2$Te$_2$Se$_2$}
\date{\today}

\begin{abstract}
	A flat band in fermionic system is a dispersionless single-particle state with a diverging effective mass and nearly zero group velocity. 
	These flat bands are expected to support exotic properties in the ground state, which might be important for a wide range of promising physical phenomena. 
	For many applications it is highly desirable to have such states in Dirac materials, but so far they have been reported only in non-magnetic Dirac systems.  
	In this work we propose a realization of topologically protected spin-polarized flat bands generated by domain walls in planar magnetic topological insulators.  
	Using first-principles material design we suggest a family of intrinsic antiferromagnetic topological insulators with an in-plane sublattice magnetization and a high N\'eel temperature.  
	Such systems can host domain walls in a natural manner.  
	For these materials, we demonstrate the existence of spin-polarized flat bands in the vicinity of the Fermi level and discuss their properties and potential applications.
\end{abstract}

\maketitle 

\section{Introduction}
The modern technology proposals require the consideration of quantum
effects, which will significantly expand the functionality of new
spintronic devices.  Of great importance is realization of such
physical phenomena as various Hall effects \cite{kane2005quantum, kane2005z, weng2015quantum}, 
a gate-tunable topological valley transport \cite{ju2015topological, hung2019direct, mania2019topological, sui2015gate, he2019silicon} and superconductivity \cite{lian2019twisted, lu2019superconductors, po2018origin, yankowitz2019tuning, cao2018unconventional}.  
In many cases, it can be
attained using specific electron states -- flat bands -- which can
arise either because of strong electronic correlations~\cite{lieb1989two, 
	mielke1991ferromagnetism,  tanaka2003stability, katsura2010ferromagnetism} 
or due to specific structural deformations~\cite{tang2014strain, Cao2018-2}.  
In absence of strongly correlated electrons, flat bands can be induced by strain
and were previously found in a number of non-magnetic materials such
as IV-VI semiconductor multilayers including topological crystalline
insulators~\cite{fogel2006direct, mironov1988superconductivity,
	tang2014strain}, and, recently, in twisted bilayer graphene
\cite{Cao2018, Cao2018-2}.  In magnetic topological materials such
states have never been reported so far.

We propose a way to generate flat bands in magnetic topological
insulators (TIs) surfaces, where massless Dirac states and the
exchange fields can serve as a platform to create spin-polarized
dispersionless states.  The latter can appear due to magnetic domain
walls (DWs) at the surface.  However, most magnetic TIs are not well
appropriated for such a realization of flat bands, either because of a
relatively low critical temperature (Curie or N\'eel temperature) or
because of strong disorder effects in magnetically doped TI's.  In
this context, a universal platform can be provided by van der Waals
(vdW) layered antiferromagnetic topological insulators (AFM
TIs)~\cite{Gong.nat2017, Huang.nat2017, otrokov2017highly,
	Otrokov.jetpl2017, hirahara2017large, klein2018probing,
	song2018giant, wang2018very}, where topological phase is governed by
the $S=\Theta T_{1/2}$ symmetry, with $\Theta$ and $T_{1/2}$ being
time-reversal and primitive-lattice translation operators,
respectively \cite{mong2010antiferromagnetic}.

As any other magnets, magnetic vdW compounds may manifest a domain
structure \cite{mcguire2015coupling}, which ensures the existence of
antiphase domain walls. Just recently Sass and coauthors have
reported on the visualization and manipulation of DWs in the
out-of-plane AFM TI MnBi$_{2-x}$Sb$_x$Te$_4$ \cite{sass2019magnetic}.
When such a DW appears in the AFM TI bulk, it can be terminated at the
sample surface. Magnetic DWs on AFM TI surface could also be induced
intentionally using the tip of a magnetic force microscope
\cite{yasuda2017quantized} or by spatially modulated external magnetic
field due to Meissner repulsion from a bulk superconductor
\cite{rosen2017chiral}, as it has been realized in Cr-doped TI
(Bi,Sb)$_2$Te$_3$. Note also that a structural step can cause the
formation of antiphase DW at rough surface of vdW AFM material in
which sublattice magnetization direction alternates along the stacking
direction.

Here, using our experience in first-principles design of topological
insulators~\cite{eremeev2012atom, hirahara2017large,
	otrokov2019prediction}, we suggest a number of AFM TI candidates
with in-plane sublattice magnetization (planar AFM TIs in the
following text) in the family of vdW systems$M$Pn$_2$Ch$_4$ ($M$=Mn,
V; Pn=Bi, Sb; Ch=Se, Te).  Remarkably, the proposed V-based compounds
have a significantly higher N\'eel temperature (in the range of 77 --
94~K) than Mn-based AFM TIs and related systems \cite{otrokov2019prediction,
	Otrokov.prl2019, klimovskikh2019variety, wu2019natural, chen2019intrinsic}.  
We verify the shift of the gapless Dirac cone in
momentum space in the compounds under study.  Employing a
tight-binding and a model Hamiltonian approaches, we demonstrate that
antiphase DW induces a bound surface state with peculiar
characteristics: it is topologically protected, dispersionless (flat)
and exhibits out-of-plane spin polarization.  Also we show that by
applying an external magnetic field perpendicular to the surface
it is possible to tune the characteristics of this DW induced state.  In
view of the unique combination of two distinctive properties, flat
bands and planar magnetic TIs, we suggest several potential
applications such as optical spin manipulations, anomalous Hall
effect and superconductive coupling between the neighboring DWs. 

\begin{figure*}[t]
	\centering
	\includegraphics{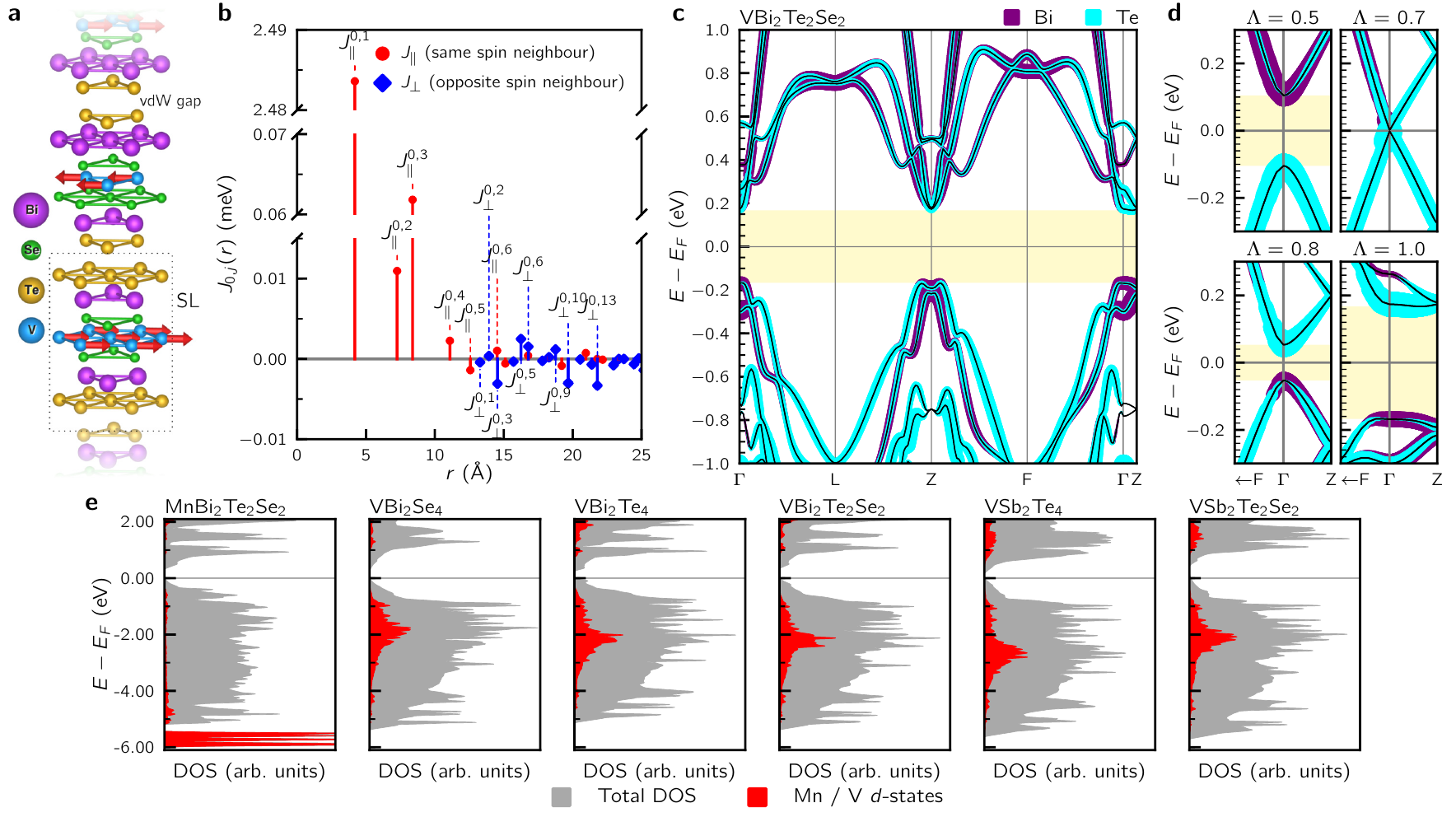}
	\caption{ \textbf{(a)} Crystal structure of bulk \VBTS{}.
		\textbf{(b)} Calculated exchange parameters $J^{0,j}$ for
		the intralayer ($J_{||}$, red circles) and interlayer
		($J_{\perp}$, blue diamonds) pair interactions as a function
		of the V--V distance $r_{0,j}$ for \VBTS{}.  \textbf{(c)}
		\VBTS{} bulk band structure. The energies are given with
		respect to the Fermi level $E_F$. The band gap is
		highlighted with light yellow. The orbital composition is
		represented by colored lines, whose thickness is
		proportional to a specific site contribution to the
		electron state.  \textbf{(d)} Bulk \VBTS{} band structure
		in the vicinity of $\Gamma$ point at different values of the
		SOC constant $\Lambda$ (SOC is not taken into account with
		$\Lambda = 0$, and fully taken into account with
		$\Lambda = 1$). The orbital composition is also present.
		\textbf{(e)} Bulk DOS of considered compounds (total and
		projected on Mn or V sites). Note, here the Fermi level is
		positioned at valence band top. }
	\label{Fig1}
\end{figure*}

\section{Results and discussion}
Based on total energy arguments we theoretically verified that the
V-based materials (\VBS{}, \VBTS{}, \VBT{}, \VSTS{}, \VST{}) and
\MBTS{} (see Table \ref{thetable}) energetically stable (see
Supplementary Table 1).  As the related compounds tend to
crystallize either in monoclinic ($C2/m$ space group) or rhombohedral
($R \bar3 m$ space group, Fig.~\ref{Fig1}a) structure
\cite{eremeev2017competing}, we compared total energies in these two
phases for each compound and found that the latter phase is
Preferable (see Supplementary Table 2).  It is important to note, that the crystal structure in
this phase can be represented by hexagonal septuple layer (SL) blocks
(e.g., Te - Bi - Se - V - Se - Bi - Te in the case of \VBTS{}),
separated by vdW gaps.  For more details of calculations see
Supplementary Note 1 and Supplementary Table 3.

\begin{table*}[t]
	\caption{\label{thetable} Differences in total energies between FM
		($E^{intra}_{FM}$), cAFM ($E^{intra}_{cAFM}$) and ncAFM
		($E^{intra}_{ncAFM}$) SL slab magnetic configurations and FM
		($E^{inter}_{FM}$) and AFM ($E^{inter}_{AFM}$) bulk magnetic
		orders.  Differences in total energies of in plane ($E_{||}$) and
		out of plane ($E_{\perp}$) spin quantization axis orientations,
		also including the dipole-dipole contribution ($E_{dip}$)).
		Magnetic moments on Mn or V sites. Calculated N\'eel temperatures
		$T_N$. Bulk band gap. $\mathbb{Z}_2$ invariant value.}
	\begin{ruledtabular}
		\begin{tabular}{ccccccc}
			Compound                     & \MBTS{}  &  \VBS{}  & \VBTS{}  &  \VBT{}  & \VSTS{}  &  \VST{}   \\ \hline\hline
			$E^{intra}_{cAFM} - E^{intra}_{FM}$ (meV/f.u.)  &  $+5.1$  &  $+9.2$  & $+13.9$  & $+16.6$  & $+11.6$  &  $+12.3$  \\
			$E^{intra}_{ncAFM} - E^{intra}_{FM}$ (meV/f.u.) &  $+6.8$  & $+17.6$  & $+25.7$  & $+29.9$  & $+14.9$  &  $+15.5$  \\
			$E^{inter}_{AFM} - E^{inter}_{FM}$ (meV/f.u.) \ & $-0.770$ & $-0.164$ & $-0.320$ & $-0.677$ & $-0.387$ & $ -0.788$ \\
			$E_{||} - E_{\perp}$ (meV/f.u.)         & $+0.053$ & $-0.092$ & $-0.311$ & $-0.176$ & $-0.004$ & $-0.087$  \\
			$E_{||} - E_{\perp} + E_{||}^{dip}$ (meV/f.u.)  & $-0.078$ & $-0.148$ & $-0.363$ & $-0.224$ & $-0.059$ & $-0.138$  \\
			Magnetic moment ($\mu_B$)            &  4.622   &  2.924   &  2.933   &  2.956   &  2.936   &   2.966   \\
			$T_N$   (K)                   &   18.6   &  80.88   &   77.1   &   78.6   &   91.6   &   93.9    \\
			Band gap (meV)                  &   256    &    55    &   334    &   233    &    11    &    125    \\
			$\mathbb{Z}_2$                  &    1     &    0     &    1     &    1     &    0     &     1
		\end{tabular}
	\end{ruledtabular}
\end{table*}

Since the interlayer magnetic coupling in similar vdW systems was
found to be rather weak compared to the intralayer one
\cite{Otrokov.prl2019, otrokov2019prediction, mcguire2015coupling}, we
consider first the magnetic order in a single SL. Total energy
calculations show ferromagnetic configuration to be preferable of the
three considered magnetic configurations, ferromagnetic (FM),
collinear antiferromagnetic (cAFM) and non-collinear antiferromagnetic
(ncAFM). Taking into account the interlayer magnetic coupling, total
energy calculations reveal the antiparallel alignment of magnetic
moments in adjacent SLs in the bulk materials (see Table
\ref{thetable}).  These results are supported by the calculated
exchange coupling parameters (Fig.~\ref{Fig1}b), which are mostly
positive for the intralayer interaction ($J^{0,i}_\parallel$) indicating
FM order, whereas the interlayer exchange parameters $J^{0,i}_\perp$ are
mostly negative, which is a distinct feature of the interlayer AFM order.

To calculate the magnetocrystalline anisotropy energy (MAE), we
consider three different spin quantization axis orientations: $[0001]$
(out-of-plane), $[10\bar{1}0]$ and $[1\bar100]$ (in-plane).  All
considered compounds were found to tend to in-plane magnetization (see
Table \ref{thetable}). \MBTS{} is the only compound which stands out,
because its in-plane magnetization is due to a strong dipole-dipole
contribution to MAE.  We did not find any significant in-plane MAE.

We note the remarkable difference between the V- and Mn-containing
compounds.  The vdW systems under consideration possess a layered
structure, where FM layers are well separated, and the interlayer
exchange coupling $J_{\perp}$ is much weaker than the intralayer one
$J_{\parallel}$.  Typically, in layered systems, the
magnetocrystalline anisotropy and the interlayer exchange coupling are
essential for establishing a magnetic order at finite temperature.  If
a two-dimensional magnet has a continuous symmetry in spin space,
there is no spontaneous magnetization at finite temperatures
\cite{viana2007anisotropy, mermin1966absence}.  For the proposed
planar AFM TIs, within the framework of our calculation accuracy, we
have not been able to identify a preferable orientation of Mn (or V)
sublattice magnetization relative to crystallographic axes in the
basal plane.  It means that these planar AFM TIs are highly
sensitive to orientational thermal fluctuations, which are expected
to hinder the establishment of an intrinsic long-range magnetic order.
This is in contrast to AFM TI \MBT{}, which is an Ising magnet with
out-of-plane easy axis.

Our calculations show that the interlayer exchange coupling in the
V-containing compounds is one or two orders of magnitude larger than
in \MBTS{} and \MBT{}~\cite{otrokov2019prediction} (see
Fig.~\ref{Fig1}b).  The underlying cause is as follows.  As follows
from the density of states (DOS) plots, Mn 3$d$-states are located far
away from the Fermi level at $\simeq -6$~eV and overlap only
marginally with \textit{p}-bands of Bi, Te and/or Se, implying the
main effect of Mn to introduce the exchange field into SL block (see
Fig.~\ref{Fig1}e).  In contrast, $3d$-states of vanadium hybridize
significantly the $p$-states of Bi/Te/Se within a wide energy range,
which provides a very strong superexchange coupling between
neighboring SLs across the vdW gap.  As a result, vanadium compounds
exhibit N\`eel temperature $ T_N $ above 75~K, which is about nearly 4
times higher than in the case of \MBTS{} (see Table \ref{thetable}).
It should be noted that predicted \MBTS{} $T_N$ is slightly lower than
in the case of out-of-plane magnetized AFM TI \MBT{} (24.3~K)
\cite{otrokov2019prediction}.  The N\'eel temperature can be roughly
estimated within the standard spin-wave theory,
$T_{N}^{SW}\sim \frac{J_{\parallel}S^{2}}{\ln(\vartheta
	\frac{J^{\parallel}}{J_{\perp}})}$ ($\vartheta$ is a model parameter
of the order of $\pi^{2}$).  Then it is clear that an increase of the
interlayer coupling by one or two orders of magnitude leads to three-
or fourfold growth of the N\'eel temperature, which is in qualitative
agreement with the calculation results presented in Table
\ref{thetable}.  Thus, the enhanced interlayer interaction due to
hybridization of the $p$-states of Bi/Te/Se and the $3d$-states of V
is crucial to stabilize a long-range AFM order at such high
temperatures.

In order to be sure that the resulting magnetic structures are stable
against relatively small lattice parameter perturbations, which may be
induced during growth process, we studied the dependence of the
magnetic structure on lattice constant values which were varied within
$\pm 3~\%$ from the equilibrium value.  Resulting equilibrium magnetic
structures were found to be insensitive to such perturbations.

All considered compounds have typical narrow-gap semiconductor band
structures with a band gap ranging from 11~meV up to 334~meV (see
Fig.~\ref{Fig1}c and Table \ref{thetable}, also see Supplementary Note 2 and Supplementary Figure 2).  
We found that the \MBTS{}, \VBTS{}, \VBT{} and \VST{}
band gaps are inverted, which is confirmed by $\mathbb{Z}_2$ invariant
calculations, which show $\mathbb{Z}_2 = 1$.  In contrast, \VBS{} and
\VSTS{} were found to have a trivial insulating bulk band structure
with $\mathbb{Z}_2 = 0$.  In order to track the band gap inversion
genesis, we calculated \VBTS{} bulk band structure at different
spin-orbit coupling (SOC) weighted with a parameter $\Lambda$ ranging
from 0 (SOC not accounted) to 1 (SOC fully accounted).  The results
clearly show that orbital composition of the band gap edges at
$\Lambda < 0.7$ is not inverted yet: highest valence band is primarily
formed by Te states, and lowest conduction band by Bi states.  At
$\Lambda \approx 0.7$ the band gap vanishes and at $\Lambda > 0.7$ it
reopens with inverted edges (see Fig.~\ref{Fig1}d).  The inversion is
caused mainly by $p_z$ states of ions close to vdW gaps (Te and Bi),
similar to tetradymite-like non-magnetic TIs.  Other topologically
nontrivial compounds under study exhibit a similar behavior.

\begin{figure*}[t]
	\centering
		\includegraphics{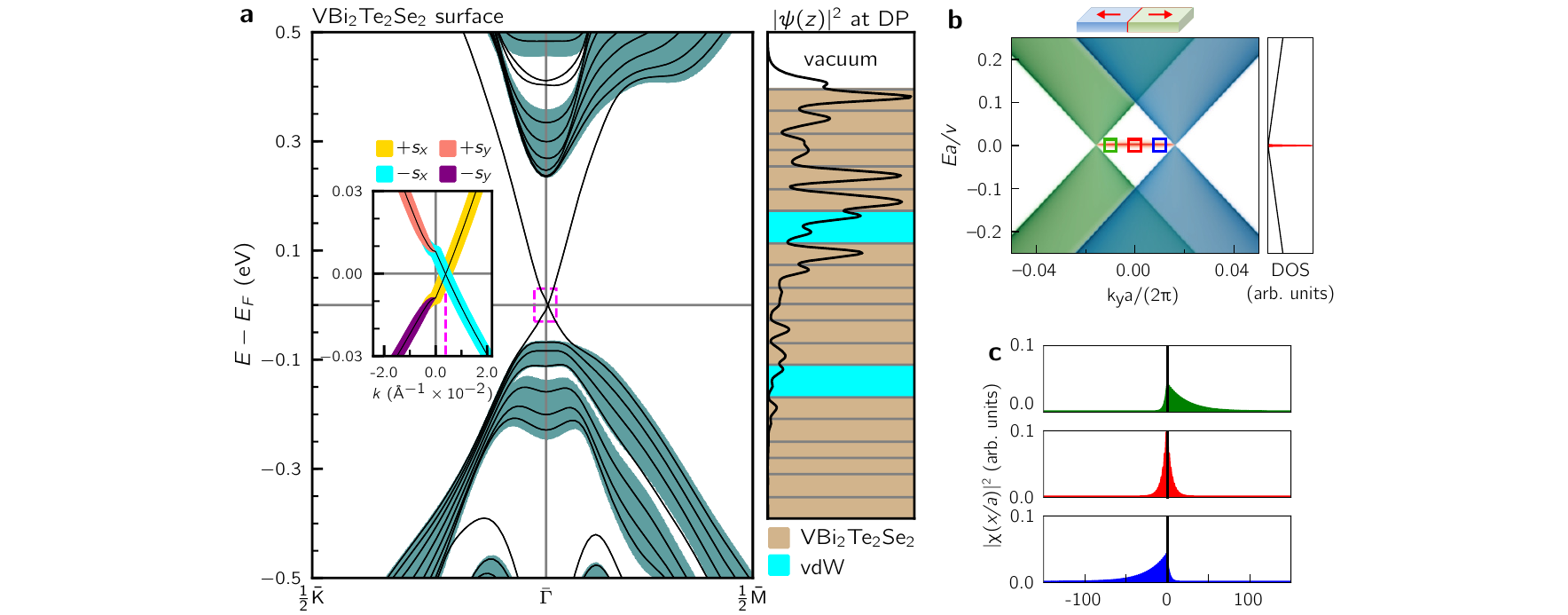}
	\caption{ \textbf{(a)} \VBTS{} surface band structure near the
		Fermi level. Sea-blue areas lines correspond to bulk bands,
		projected on two-dimensional Brillouin zone (2D BZ), and the
		black lines to surface bands. Surface spin texture (inset on
		the left) is represented by color lines, which thickness is
		proportional to spin projection value. The right panel
		depicts $|\psi(z)|^2$ at Dirac point (DP), with $\psi$ being
		one-electron wavefunction.  \textbf{(b)} Spectral density of
		the electron states on the surface containing single
		antiphase DW . For generality, the scales of the axes are
		presented in dimensionless units constructed by combination
		of energy and momentum with model parameters. The spectral
		density corresponding to the left (right) semi-infinite
		region is represented by blue (green) color,and flat band by
		red color. DOS at the $\bar{\Gamma}$ point of 2D BZ is shown
		on the right side of the panel. Black solid line represents
		DOS for the Dirac cones, the red peak is DOS for the flat
		band. The magnetic configuration is schematically
		illustrated on the top so that red arrows in blue and green
		regions indicate magnetization directions in the vicinity of
		DW.  \textbf{(c)} The charge density distribution of
		zero-energy bound states as a function of distance from the
		DW. The energy-momenta values are marked by color squares in
		the panel \textbf{(b)}. }
	\label{Fig2}
\end{figure*}

The inversion ensures the appearance of a topological surface state
(TSS) on the (0001) surface of \MBTS{}, \VBTS{}, \VBT{} and \VST{}.
As an example, Fig.~\ref{Fig2}a presents the spectrum of such surface
state of\VBTS{}.  As may be seen from the inset, the
Dirac point is slightly shifted from the $\bar \Gamma$.  The
direction of the shift is normal to the magnetization $\mathbf{M}$ of
the topmost SL, and its value is proportional to its magnitude
$|\mathbf{M}|$. Surface band structures of the other topologically
non-trivial compounds can be found in Supplementary Note 2 and Supplementary Figure 3.  Thus,
our DFT calculations demonstrate that all the topologically nontrivial
compounds under study are characterized by a gapless Dirac state with
a helical spin texture on the ideal (0001) surface.

\begin{figure*}[t]
	\centering
		\includegraphics{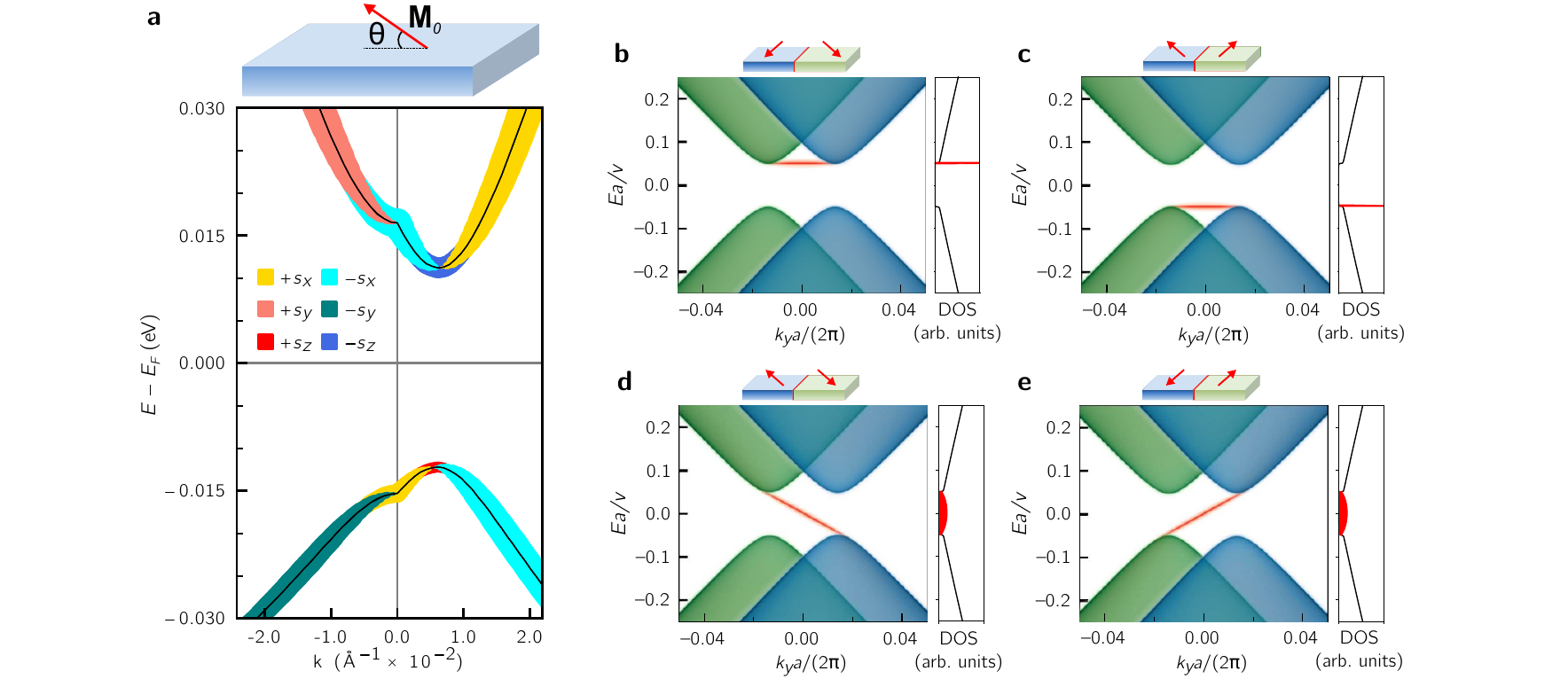}
	\caption{\textbf{(a)} \MBTS{} surface band structure near the Fermi level in the case of mixed magnetization. 
		\textbf{(b, c, d, e)} The same as Fig.~\ref{Fig2}b but with additional negative \textbf{(b)},   positive \textbf{(c)} and  opposite \textbf{(d, e)} out-of-plane magnetization components. }
	\label{Fig3}
\end{figure*}

The right panel in Fig.~\ref{Fig2}a indicates that the Dirac fermions
are predominantly localized within the topmost SL.  Their behavior,
affected by the exchange field, $\sim\mathbf{M}$, originated from the
local moments of the same SL, can be described by an effective
two-dimensional Hamiltonian
\begin{equation}\label{Eq_1}
H(\mathbf{k})=\upsilon(k_{y}\sigma_{x}-k_{x}\sigma_{y})-j(M_{x}\sigma_{x}+
M_{y}\sigma_{y})-j_{\perp}\sigma_{z}M_{z},
\end{equation}
where $\upsilon$ is the Fermi velocity, $\mathbf{k}=(k_{x},k_{y})$ is
the in-plane momentum, $\sigma_{x}$, $\sigma_{y}$ and $\sigma_{z}$ are
the Pauli matrices acting in spin space.  For the definiteness, we
assign $j,j_{\perp}>0$ and $\upsilon >0$.  The first term captures the
presence of the Dirac-like quasiparticles with a linear spectrum and a
perfect spin-momentum locking.  The effective exchange integrals $j$
and $j_{\perp}$ couple the surface quasiparticle spins with the local
magnetization of the uppermost SL, $\mathbf{M} = (M_{x},M_{y},M_{z})$,
which can generically be in an arbitrary direction.
In the case of the spatially homogeneous magnetization $\mathbf{M}$,
the energy spectrum of the Hamiltonian~(\ref{Eq_1}) is given by the
relation
$E^{2}(\mathbf{k})=(\upsilon k_{x}+jM_{y})^{2}+(\upsilon
k_{y}-jM_{x})^{2}+j_{\perp}^{2}M_{z}^{2}$.

To explore how the spatially inhomogeneous magnetization
$\mathbf{M}(x,y)$ affects the surface state of AFM TI, we use a model
of a single rigid DW.  In Hamiltonian~(\ref{Eq_1}) the $(x,y)$ plane
is assumed to be divided in two semi-infinite uniformly ordered
regions with oposite polarizations so that $\mathbf{M}(x,y)$ changes
its direction upon crossing the linear boundary $x=0$, but keeps its
magnitude.  We specify the space profile of the DW in the form
$\mathbf{M}(x,y)=M_{0}(\sin\theta h(x),0,\cos\theta)$, where $h(x)$ is
the Heaviside function, $|\mathbf{M}(x,y)|= M_{0}= const$.  Here the
magnetization vector can rotate through an angle $\pi-\theta$ out of
the plane.  Due to the periodicity along $\mathbf{y}$ direction, the
momentum $k_{y}$ is a good quantum number.  To describe the fermion
state hosted by the DW we apply the analytical approach as well as the
numerical tight-binding calculations.

In following, we address the planar AFM TI surface where the local
moments lie in the plane and have opposite directions in the right and
left domains, i.e.  $\mathbf{M}(x,y)=M_{0}(h(x),0,0)$ (see
Fig.~\ref{Fig2}b).
The surface with such "tail-to-tail" DW harbors two types of
quasiparticles manifesting utterly distinct behaviors: on the one
hand, the 2D \textit{massless} Dirac fermions; on the other hand, the
1D heavy fermions with \textit{infinitely large effective mass}.  The
pair of the Dirac cones, shifted to momenta $\pm k_{0}$, where
$k_{0}= \frac{jM_{0}}{\upsilon}$, with respect to the Brillouin zone
center, corresponds to the two semi-infinite domains with opposite
magnetization.  The flat band exists within the interval between the
Dirac nodes, $|k_{y}|<k_{0}$.  Remarkably, this particular state does
not disperse in $k_{y}$ at zero energy, $E(\mathbf{k})=0$, forming a
sharp peak in DOS at $E=0$ against the linear dependence of Dirac
fermions DOS (see right panel on Fig.~\ref{Fig2}b).  This flat band
state is topologically protected, which is originated from the Berry
phase of $\pi$ for each Dirac node, and, therefore, can not be
destroyed by DW imperfections. The probability density of the flat
band decays exponentially away from the DW on the scale
$|\chi(x)|^{2}\sim\exp(-2k_{0}|x|)$ as demonstrated in
Fig.~\ref{Fig2}c.  Furthermore, the dispersionless state is fully spin
polarized.  Therefore, the expectation value of the spin angular
momentum is zero for the in-plane components,
$\langle \sigma_{x}\rangle=\langle \sigma_{x}\rangle=0$, but it is
non-zero in the direction normal to the surface, i.e., normal to the
easy plane, $\langle \sigma_{z} \rangle \neq0$.  Note, the spin
polarization of the flat band induced by the "head-to-head" DW is
antiparallel to the one induced by the "tail-to-tail" DW.

In general, the surface magnetization can not be attached tightly to
the plane.  For example, Mn or V sublattice magnetization of a planar
AFM TI can acquire non-zero out-of-plane component $M_{z}$ due to an
external magnetic field or magnetic proximity effect.  In the case of
an AFM TI surface hosting an isolated DW with the spatial profile
$\mathbf{M}(x,y)= M_{0}(\sin\theta h(x),0,\cos\theta)$, the uniform
out-of-plane component $M_{z}$ breaks the spin degeneracy opening the
energy gaps at the Dirac points $\pm k_{0}\sin\theta$ in the spectrum
(see Fig.~\ref{Fig3}b,c).  As follows from the figures, "tail-to-tail"
DW creates the dispersionless state with the energy $E=j_{\perp}M_{z}$
($E=-j_{\perp}M_{z}$), which connects the band edges of the two gapped
cones with dispersion
$E^{2}(\mathbf{k}) = (\upsilon k_{x})^{2} + \upsilon^{2} (k_{y}\pm
k_{0}\sin\theta)^{2} + j_{\perp}^{2}M_{z}^{2}$.  Correspondingly, the
keen peak in the density of states appears just at the band
edge. Indeed we observe such a gap opening within \textit{ab-initio}
calculations when the perfect planar AFM phase is subjected to an
out-of-plane Zeeman perturbation.  The latter contributes to the gap
opening at the Dirac point in the surface state spectrum of \MBTS{}
(see Fig.~\ref{Fig3}a).

According to the numerical simulations, \MBTS{} is vdW AFM TI with
in-plane anisotropy, whereas \MBT{} has been identified as an
out-of-plane AFM TI.  Therefore, it is natural to assume that the
solid solution MnBi$_2$(Te$_{1-x}$Se$_x$)$_4$ at the certain value
$ 0 < x < \frac{1}{2} $ would have an AFM order with the sublattice
magnetization directed to the angle $0 < |\theta | < \frac{\pi}{2}$ to
the basal plane, keeping nontrivial invariant $\mathbb{Z}_2 \neq 0$.
In our approach, in the presence of the antiphase DW, the surface
magnetization of such a material is modelled with the spatial profile
$\mathbf{M}(x,y) = M_0(\sin{\theta h(x)}, 0, \cos{\theta h(x)})$.
Interestingly, the DW induced bound state is associated with the
linear spectral branch,
$ E(k) = \pm \frac{j_\perp}{j} \cot{\theta v k_y} $, which spans the
magnetic gap, $ 2|j_\perp M_0 \cos{\theta}| $, and connects edges of
the bands originated from the opposite magnetic domains (see
Fig.~\ref{Fig3}d, e).  Thus, the surface band structure of vdW AFM TI
materials can be tuned by different factors (such as chemical
composition, strain, etc.) to change the properties of the DW bound
state from almost a dispersionless band (with huge mass) to a massless
band near the Fermi level.

\section{Conclusion}
In this paper we proposed a tetradymite-like planar AFM TI family and
by means of \textit{ab initio} calculations we determined their
equilibrium crystal, electronic and magnetic structure.  We found all
considered compounds to be layered antiferromagnets with an in-plane
magnetization.

The proposed V-based compounds have N\'eel temperature in the range of
77 -- 94 K, which is significantly higher than in the case of Mn-based
AFM TIs.  We showed that the critical temperature can strongly depend
on the chemical composition, in particular, Sb--containing compounds
show increasing N\'eel temperature of almost 20 K with respect to Bi
ones.  All the considered compounds possess a typical semiconductor
bulk band structure.  We found bulk \MBTS{}, \VBTS{}, \VBT{} and
\VST{} to exhibit $\mathbb{Z}_2 = 1$ and an AFM TI phase.  Such
compounds are characterized by a gapless surface state on the (0001)
surface, which has a helical spin texture similar to an nonmagnetic
TI. The topmost SL magnetization shifts Dirac point from $\bar \Gamma$.

We demonstrated that magnetic inhomogeneities likes DWs on the (0001)
surface can prompt the appearance of topological one-dimensional flat
bands, which give rise to a sharp peak in the DOS near the Fermi
level.  We showed that flat band states can be effectively tuned by
applying an external magnetic field perpendicular to the surface
plane.  In this context, AFM TIs with the in-plane sublattice
magnetization provide a very special and rich platform to study
surface electronic properties.

The appearance of flat bands with fully spin-polarized electron states
leads to some unusual effects in these materials. For example, the
optical excitation of electrons from the flat band can lead to an
observable spin-resolved photoelectric effect, e.g. spin- and valley-polarized
currents.

The flat band state can manifest itself in the anomalous Hall effect on a single DW. Indeed, free electrons in the uppermost SL, transmitted through the DW, are subjected to the out-of-plane polarisation associated with the state. Due to SOC it can lead to the transverse current.

Since the Fermi energy is pinned at the flat band energy, we can
also predict the appearance of superconductivity related to the
coupling of heavy electrons from the neighboring DWs.  Indeed, the
electrons from neighboring DWs have opposite spin directions, and the
interaction between them via phonons can be rather strong since the
electron localization allows to release the momentum conservation
condition in the electron-phonon interaction. It should be also noted
that the intra-DW electron-electron repulsion does not affect the
superconductivity related to pairing at different
DWs\cite{efetov1975pairing}. Finally, we can expect an enhancement of
the critical temperature for superconductivity transition thanks to an
infinite electron density of states.

\appendix*
\section{Methods} 
Bulk crystal structures, magnetic order, MAE, bulk and surface band
structures were investigated using the projector augmented-wave method
(PAW) \cite{blochl1994projector} implemented in VASP package
\cite{kresse1993ab, kresse1996efficient, kresse1996g}.
Exchange-correlation effects were taken into account using
Perdew-Burke-Ernzerhof generalized gradient approximation (GGA)
\cite{perdew1996generalized}.  Spin-orbit coupling was treated using
the second variation technique \cite{koelling1977technique}.  DFT-D3
method \cite{grimme2010consistent} was used to accurately describe the
van der Waals interaction.  The plane wave energy cutoff was chosen
exclusively for each compound (280~eV for \MBTS{}, 240~eV for \VBS{},
\VBTS{} and \VBT{}, 275~eV for \VSTS{} and 250~eV for \VST{}) and was
kept constant through all calculations.  The energy convergence
criterion was set to $10^{-6}$~eV for all types of calculations except
magnetocrystalline anisotropy study, for which it was decreased down
to $10^{-7}$~eV.  FM phases were modeled using a rhombohedral cell
containing one Mn or V atom (1 f.u.) and monoclinic cell containing 4
Mn or V atoms (4 f.u.), respectively.  AFM bulk phases was modeled
using rhombohedral cell, containing 2 Mn atoms (2 f.u.) and hexagonal
cell, containing 6 Mn atoms (6 f.u.).  Collinear AFM and non-collinear
AFM phases were modeled using rectangular $(1~\times~\sqrt{3})$ and
$(\sqrt{3}~\times~\sqrt{3})R30\degree$ cell, respectively.  All
ferromagnetic slabs were studied using convenient hexagonal cell.
Magnetocrystalline anisotropy studies were performed on the same
hexagonal cell.  Surfaces were modeled within repeating slabs model.

Mn and V $3d$-states were treated using a GGA$+U$ approach
\cite{anisimov1991band, dudarev1998electron}.  The values of
$U_{\mathrm{eff}}$ were calculated using a linear response technique
proposed by Cococcioni et al.~\cite{cococcioni2005linear} Adopted $U$
values were $5.3$, $4.8$, $5.0$, $4.7$, $4.6$ and $5.0$~eV for
\MBTS{}, \VBS{}, \VBTS{}, \VBT{}, \VSTS{} and \VST{} respectively.
Also we looked into other close $U_{\mathrm{eff}}$ values and we did
not find any qualitative changes in calculation results.

Calculations of $\mathbb{Z}_2$ invariants were performed using Z2Pack
\cite{soluyanov2011computing, gresch2017z2pack, marzari1997maximally,
	mostofi2008wannier90}.

In order to obtain exchange coupling parameters, we used the magnetic
force theorem\cite{liechtenstein1987local} as it is implemented within
the multiple scattering theory package Hutsepot
\cite{Hoffmann2020pssb}, along with the full charge density
approximation \cite{vitos1994full}.  They were confirmed by
Monte-Carlo simulations based on the classical Heisenberg Hamiltonian
with the obtained exchange coupling parameters from above.  The heat
capacity was used as indication for the magnetic phase transition.
Monte Carlo results were checked for convergence of all simulation
parameters, i.e., simulation size and Monte Carlo steps.  More
technical details can be found in Ref. \cite{Hoffmann2020pssb}.

The introduced toy model, Eq. (1), may be directly implemented to
analytically describe low-energy fermions at the surface of planar AFM
TI for various inhomogeneous magnetization configurations in uppermost
SL. This consideration is restricted to simple configurations (in the
form of rigid one-dimensional DWs) which allowed us to find the exact
solution for the corresponding eigen-state problem. Indeed, we have
obtained a modification of the energy spectrum and the envelope wave
function spatial profile of the surface states caused by a magnetic DW
presence. These results are consistent with those of tight-binding
study of the model regularized on square lattice.

\begin{acknowledgments}
	We are thankful for support from Tomsk State University
	competitiveness improvement programme (project 8.1.01.2018) and from
	Saint Petersburg State University (Grant ID 40990069).
	E.K.P. acknowledges support from RFBR within the research projects
	No.~18-32-00728 (the study of V-based compounds crystal, magnetic
	and band structures) and No.~19-32-90250 (the study of V-based
	compounds critical temperatures).  V.N.M., T.V.M. and
	I.P.R. acknowledge support from RSF within the research project
	No.~18-12-00169 (the study DWs by model Hamiltonian and
	tight-binding approaches).  A.E. acknowledges support from DFG
	through priority program SPP1666 (Topological Insulators) and OeAD
	Grants No.~HR 07/2018 and No.~PL 03/2018.  V.K.D. acknowledges
	support from the National Science Center of Poland under the project
	No.~DEC-2017/27/B/ST3/02881.  E.V.C. acknowledges support from RFBR
	within the research project No.~18-52-06009.  Calculations performed
	at the Research Park of St.-Petersburg State University ``Computing
	Center'' and SKIF-Cyberia supercomputer of National Research Tomsk
	State University.
\end{acknowledgments}

\vfill
\section*{Author contributions}
First-principles calculations were performed by E.K.P. and A.E. 
Monte-Carlo simulations were done by M.H.
Model Hamiltonian approach was developed by V.N.M. 
Tight-binding calculations were performed by I.P.R.
Figures were produced by E.K.P., T.V.M. and I.P.R. 
Project planning was done by E.V.C., E.K.P., T.V.M., I.P.R., V.N.M. and A.E. 
All authors contributed to the discussion and to writing the manuscript.

\bibliography{inplane_AFMTI_bib}

\begin{thebibliography}{62}%
\makeatletter
\providecommand \@ifxundefined [1]{%
 \@ifx{#1\undefined}
}%
\providecommand \@ifnum [1]{%
 \ifnum #1\expandafter \@firstoftwo
 \else \expandafter \@secondoftwo
 \fi
}%
\providecommand \@ifx [1]{%
 \ifx #1\expandafter \@firstoftwo
 \else \expandafter \@secondoftwo
 \fi
}%
\providecommand \natexlab [1]{#1}%
\providecommand \enquote  [1]{``#1''}%
\providecommand \bibnamefont  [1]{#1}%
\providecommand \bibfnamefont [1]{#1}%
\providecommand \citenamefont [1]{#1}%
\providecommand \href@noop [0]{\@secondoftwo}%
\providecommand \href [0]{\begingroup \@sanitize@url \@href}%
\providecommand \@href[1]{\@@startlink{#1}\@@href}%
\providecommand \@@href[1]{\endgroup#1\@@endlink}%
\providecommand \@sanitize@url [0]{\catcode `\\12\catcode `\$12\catcode
  `\&12\catcode `\#12\catcode `\^12\catcode `\_12\catcode `\%12\relax}%
\providecommand \@@startlink[1]{}%
\providecommand \@@endlink[0]{}%
\providecommand \url  [0]{\begingroup\@sanitize@url \@url }%
\providecommand \@url [1]{\endgroup\@href {#1}{\urlprefix }}%
\providecommand \urlprefix  [0]{URL }%
\providecommand \Eprint [0]{\href }%
\providecommand \doibase [0]{http://dx.doi.org/}%
\providecommand \selectlanguage [0]{\@gobble}%
\providecommand \bibinfo  [0]{\@secondoftwo}%
\providecommand \bibfield  [0]{\@secondoftwo}%
\providecommand \translation [1]{[#1]}%
\providecommand \BibitemOpen [0]{}%
\providecommand \bibitemStop [0]{}%
\providecommand \bibitemNoStop [0]{.\EOS\space}%
\providecommand \EOS [0]{\spacefactor3000\relax}%
\providecommand \BibitemShut  [1]{\csname bibitem#1\endcsname}%
\let\auto@bib@innerbib\@empty
\bibitem [{\citenamefont {Kane}\ and\ \citenamefont
  {Mele}(2005{\natexlab{a}})}]{kane2005quantum}%
  \BibitemOpen
  \bibfield  {author} {\bibinfo {author} {\bibfnamefont {C.~L.}\ \bibnamefont
  {Kane}}\ and\ \bibinfo {author} {\bibfnamefont {E.~J.}\ \bibnamefont
  {Mele}},\ }\href@noop {} {\bibfield  {journal} {\bibinfo  {journal} {Physical
  review letters}\ }\textbf {\bibinfo {volume} {95}},\ \bibinfo {pages}
  {226801} (\bibinfo {year} {2005}{\natexlab{a}})}\BibitemShut {NoStop}%
\bibitem [{\citenamefont {Kane}\ and\ \citenamefont
  {Mele}(2005{\natexlab{b}})}]{kane2005z}%
  \BibitemOpen
  \bibfield  {author} {\bibinfo {author} {\bibfnamefont {C.~L.}\ \bibnamefont
  {Kane}}\ and\ \bibinfo {author} {\bibfnamefont {E.~J.}\ \bibnamefont
  {Mele}},\ }\href@noop {} {\bibfield  {journal} {\bibinfo  {journal} {Physical
  review letters}\ }\textbf {\bibinfo {volume} {95}},\ \bibinfo {pages}
  {146802} (\bibinfo {year} {2005}{\natexlab{b}})}\BibitemShut {NoStop}%
\bibitem [{\citenamefont {Weng}\ \emph {et~al.}(2015)\citenamefont {Weng},
  \citenamefont {Yu}, \citenamefont {Hu}, \citenamefont {Dai},\ and\
  \citenamefont {Fang}}]{weng2015quantum}%
  \BibitemOpen
  \bibfield  {author} {\bibinfo {author} {\bibfnamefont {H.}~\bibnamefont
  {Weng}}, \bibinfo {author} {\bibfnamefont {R.}~\bibnamefont {Yu}}, \bibinfo
  {author} {\bibfnamefont {X.}~\bibnamefont {Hu}}, \bibinfo {author}
  {\bibfnamefont {X.}~\bibnamefont {Dai}}, \ and\ \bibinfo {author}
  {\bibfnamefont {Z.}~\bibnamefont {Fang}},\ }\href@noop {} {\bibfield
  {journal} {\bibinfo  {journal} {Advances in Physics}\ }\textbf {\bibinfo
  {volume} {64}},\ \bibinfo {pages} {227} (\bibinfo {year} {2015})}\BibitemShut
  {NoStop}%
\bibitem [{\citenamefont {Ju}\ \emph {et~al.}(2015)\citenamefont {Ju},
  \citenamefont {Shi}, \citenamefont {Nair}, \citenamefont {Lv}, \citenamefont
  {Jin}, \citenamefont {Velasco~Jr}, \citenamefont {Ojeda-Aristizabal},
  \citenamefont {Bechtel}, \citenamefont {Martin}, \citenamefont {Zettl} \emph
  {et~al.}}]{ju2015topological}%
  \BibitemOpen
  \bibfield  {author} {\bibinfo {author} {\bibfnamefont {L.}~\bibnamefont
  {Ju}}, \bibinfo {author} {\bibfnamefont {Z.}~\bibnamefont {Shi}}, \bibinfo
  {author} {\bibfnamefont {N.}~\bibnamefont {Nair}}, \bibinfo {author}
  {\bibfnamefont {Y.}~\bibnamefont {Lv}}, \bibinfo {author} {\bibfnamefont
  {C.}~\bibnamefont {Jin}}, \bibinfo {author} {\bibfnamefont {J.}~\bibnamefont
  {Velasco~Jr}}, \bibinfo {author} {\bibfnamefont {C.}~\bibnamefont
  {Ojeda-Aristizabal}}, \bibinfo {author} {\bibfnamefont {H.~A.}\ \bibnamefont
  {Bechtel}}, \bibinfo {author} {\bibfnamefont {M.~C.}\ \bibnamefont {Martin}},
  \bibinfo {author} {\bibfnamefont {A.}~\bibnamefont {Zettl}},  \emph
  {et~al.},\ }\href@noop {} {\bibfield  {journal} {\bibinfo  {journal}
  {Nature}\ }\textbf {\bibinfo {volume} {520}},\ \bibinfo {pages} {650}
  (\bibinfo {year} {2015})}\BibitemShut {NoStop}%
\bibitem [{\citenamefont {Hung}\ \emph {et~al.}(2019)\citenamefont {Hung},
  \citenamefont {Camsari}, \citenamefont {Zhang}, \citenamefont {Upadhyaya},\
  and\ \citenamefont {Chen}}]{hung2019direct}%
  \BibitemOpen
  \bibfield  {author} {\bibinfo {author} {\bibfnamefont {T.~Y.}\ \bibnamefont
  {Hung}}, \bibinfo {author} {\bibfnamefont {K.~Y.}\ \bibnamefont {Camsari}},
  \bibinfo {author} {\bibfnamefont {S.}~\bibnamefont {Zhang}}, \bibinfo
  {author} {\bibfnamefont {P.}~\bibnamefont {Upadhyaya}}, \ and\ \bibinfo
  {author} {\bibfnamefont {Z.}~\bibnamefont {Chen}},\ }\href@noop {} {\bibfield
   {journal} {\bibinfo  {journal} {Science advances}\ }\textbf {\bibinfo
  {volume} {5}},\ \bibinfo {pages} {eaau6478} (\bibinfo {year}
  {2019})}\BibitemShut {NoStop}%
\bibitem [{\citenamefont {Mania}\ \emph {et~al.}(2019)\citenamefont {Mania},
  \citenamefont {Cadore}, \citenamefont {Taniguchi}, \citenamefont {Watanabe},\
  and\ \citenamefont {Campos}}]{mania2019topological}%
  \BibitemOpen
  \bibfield  {author} {\bibinfo {author} {\bibfnamefont {E.}~\bibnamefont
  {Mania}}, \bibinfo {author} {\bibfnamefont {A.}~\bibnamefont {Cadore}},
  \bibinfo {author} {\bibfnamefont {T.}~\bibnamefont {Taniguchi}}, \bibinfo
  {author} {\bibfnamefont {K.}~\bibnamefont {Watanabe}}, \ and\ \bibinfo
  {author} {\bibfnamefont {L.}~\bibnamefont {Campos}},\ }\href@noop {}
  {\bibfield  {journal} {\bibinfo  {journal} {Communications Physics}\ }\textbf
  {\bibinfo {volume} {2}},\ \bibinfo {pages} {6} (\bibinfo {year}
  {2019})}\BibitemShut {NoStop}%
\bibitem [{\citenamefont {Sui}\ \emph {et~al.}(2015)\citenamefont {Sui},
  \citenamefont {Chen}, \citenamefont {Ma}, \citenamefont {Shan}, \citenamefont
  {Tian}, \citenamefont {Watanabe}, \citenamefont {Taniguchi}, \citenamefont
  {Jin}, \citenamefont {Yao}, \citenamefont {Xiao} \emph
  {et~al.}}]{sui2015gate}%
  \BibitemOpen
  \bibfield  {author} {\bibinfo {author} {\bibfnamefont {M.}~\bibnamefont
  {Sui}}, \bibinfo {author} {\bibfnamefont {G.}~\bibnamefont {Chen}}, \bibinfo
  {author} {\bibfnamefont {L.}~\bibnamefont {Ma}}, \bibinfo {author}
  {\bibfnamefont {W.-Y.}\ \bibnamefont {Shan}}, \bibinfo {author}
  {\bibfnamefont {D.}~\bibnamefont {Tian}}, \bibinfo {author} {\bibfnamefont
  {K.}~\bibnamefont {Watanabe}}, \bibinfo {author} {\bibfnamefont
  {T.}~\bibnamefont {Taniguchi}}, \bibinfo {author} {\bibfnamefont
  {X.}~\bibnamefont {Jin}}, \bibinfo {author} {\bibfnamefont {W.}~\bibnamefont
  {Yao}}, \bibinfo {author} {\bibfnamefont {D.}~\bibnamefont {Xiao}},  \emph
  {et~al.},\ }\href@noop {} {\bibfield  {journal} {\bibinfo  {journal} {Nature
  Physics}\ }\textbf {\bibinfo {volume} {11}},\ \bibinfo {pages} {1027}
  (\bibinfo {year} {2015})}\BibitemShut {NoStop}%
\bibitem [{\citenamefont {He}\ \emph {et~al.}(2019)\citenamefont {He},
  \citenamefont {Liang}, \citenamefont {Yuan}, \citenamefont {Qiu},
  \citenamefont {Chen}, \citenamefont {Zhao},\ and\ \citenamefont
  {Dong}}]{he2019silicon}%
  \BibitemOpen
  \bibfield  {author} {\bibinfo {author} {\bibfnamefont {X.-T.}\ \bibnamefont
  {He}}, \bibinfo {author} {\bibfnamefont {E.-T.}\ \bibnamefont {Liang}},
  \bibinfo {author} {\bibfnamefont {J.-J.}\ \bibnamefont {Yuan}}, \bibinfo
  {author} {\bibfnamefont {H.-Y.}\ \bibnamefont {Qiu}}, \bibinfo {author}
  {\bibfnamefont {X.-D.}\ \bibnamefont {Chen}}, \bibinfo {author}
  {\bibfnamefont {F.-L.}\ \bibnamefont {Zhao}}, \ and\ \bibinfo {author}
  {\bibfnamefont {J.-W.}\ \bibnamefont {Dong}},\ }\href@noop {} {\bibfield
  {journal} {\bibinfo  {journal} {Nature communications}\ }\textbf {\bibinfo
  {volume} {10}},\ \bibinfo {pages} {872} (\bibinfo {year} {2019})}\BibitemShut
  {NoStop}%
\bibitem [{\citenamefont {Lian}\ \emph {et~al.}(2019)\citenamefont {Lian},
  \citenamefont {Wang},\ and\ \citenamefont {Bernevig}}]{lian2019twisted}%
  \BibitemOpen
  \bibfield  {author} {\bibinfo {author} {\bibfnamefont {B.}~\bibnamefont
  {Lian}}, \bibinfo {author} {\bibfnamefont {Z.}~\bibnamefont {Wang}}, \ and\
  \bibinfo {author} {\bibfnamefont {B.~A.}\ \bibnamefont {Bernevig}},\
  }\href@noop {} {\bibfield  {journal} {\bibinfo  {journal} {Physical review
  letters}\ }\textbf {\bibinfo {volume} {122}},\ \bibinfo {pages} {257002}
  (\bibinfo {year} {2019})}\BibitemShut {NoStop}%
\bibitem [{\citenamefont {Lu}\ \emph {et~al.}(2019)\citenamefont {Lu},
  \citenamefont {Stepanov}, \citenamefont {Yang}, \citenamefont {Xie},
  \citenamefont {Aamir}, \citenamefont {Das}, \citenamefont {Urgell},
  \citenamefont {Watanabe}, \citenamefont {Taniguchi}, \citenamefont {Zhang}
  \emph {et~al.}}]{lu2019superconductors}%
  \BibitemOpen
  \bibfield  {author} {\bibinfo {author} {\bibfnamefont {X.}~\bibnamefont
  {Lu}}, \bibinfo {author} {\bibfnamefont {P.}~\bibnamefont {Stepanov}},
  \bibinfo {author} {\bibfnamefont {W.}~\bibnamefont {Yang}}, \bibinfo {author}
  {\bibfnamefont {M.}~\bibnamefont {Xie}}, \bibinfo {author} {\bibfnamefont
  {M.~A.}\ \bibnamefont {Aamir}}, \bibinfo {author} {\bibfnamefont
  {I.}~\bibnamefont {Das}}, \bibinfo {author} {\bibfnamefont {C.}~\bibnamefont
  {Urgell}}, \bibinfo {author} {\bibfnamefont {K.}~\bibnamefont {Watanabe}},
  \bibinfo {author} {\bibfnamefont {T.}~\bibnamefont {Taniguchi}}, \bibinfo
  {author} {\bibfnamefont {G.}~\bibnamefont {Zhang}},  \emph {et~al.},\
  }\href@noop {} {\bibfield  {journal} {\bibinfo  {journal} {arXiv preprint
  arXiv:1903.06513}\ } (\bibinfo {year} {2019})}\BibitemShut {NoStop}%
\bibitem [{\citenamefont {Po}\ \emph {et~al.}(2018)\citenamefont {Po},
  \citenamefont {Zou}, \citenamefont {Vishwanath},\ and\ \citenamefont
  {Senthil}}]{po2018origin}%
  \BibitemOpen
  \bibfield  {author} {\bibinfo {author} {\bibfnamefont {H.~C.}\ \bibnamefont
  {Po}}, \bibinfo {author} {\bibfnamefont {L.}~\bibnamefont {Zou}}, \bibinfo
  {author} {\bibfnamefont {A.}~\bibnamefont {Vishwanath}}, \ and\ \bibinfo
  {author} {\bibfnamefont {T.}~\bibnamefont {Senthil}},\ }\href@noop {}
  {\bibfield  {journal} {\bibinfo  {journal} {Physical Review X}\ }\textbf
  {\bibinfo {volume} {8}},\ \bibinfo {pages} {031089} (\bibinfo {year}
  {2018})}\BibitemShut {NoStop}%
\bibitem [{\citenamefont {Yankowitz}\ \emph {et~al.}(2019)\citenamefont
  {Yankowitz}, \citenamefont {Chen}, \citenamefont {Polshyn}, \citenamefont
  {Zhang}, \citenamefont {Watanabe}, \citenamefont {Taniguchi}, \citenamefont
  {Graf}, \citenamefont {Young},\ and\ \citenamefont
  {Dean}}]{yankowitz2019tuning}%
  \BibitemOpen
  \bibfield  {author} {\bibinfo {author} {\bibfnamefont {M.}~\bibnamefont
  {Yankowitz}}, \bibinfo {author} {\bibfnamefont {S.}~\bibnamefont {Chen}},
  \bibinfo {author} {\bibfnamefont {H.}~\bibnamefont {Polshyn}}, \bibinfo
  {author} {\bibfnamefont {Y.}~\bibnamefont {Zhang}}, \bibinfo {author}
  {\bibfnamefont {K.}~\bibnamefont {Watanabe}}, \bibinfo {author}
  {\bibfnamefont {T.}~\bibnamefont {Taniguchi}}, \bibinfo {author}
  {\bibfnamefont {D.}~\bibnamefont {Graf}}, \bibinfo {author} {\bibfnamefont
  {A.~F.}\ \bibnamefont {Young}}, \ and\ \bibinfo {author} {\bibfnamefont
  {C.~R.}\ \bibnamefont {Dean}},\ }\href@noop {} {\bibfield  {journal}
  {\bibinfo  {journal} {Science}\ }\textbf {\bibinfo {volume} {363}},\ \bibinfo
  {pages} {1059} (\bibinfo {year} {2019})}\BibitemShut {NoStop}%
\bibitem [{\citenamefont {Cao}\ \emph {et~al.}(2018{\natexlab{a}})\citenamefont
  {Cao}, \citenamefont {Fatemi}, \citenamefont {Fang}, \citenamefont
  {Watanabe}, \citenamefont {Taniguchi}, \citenamefont {Kaxiras},\ and\
  \citenamefont {Jarillo-Herrero}}]{cao2018unconventional}%
  \BibitemOpen
  \bibfield  {author} {\bibinfo {author} {\bibfnamefont {Y.}~\bibnamefont
  {Cao}}, \bibinfo {author} {\bibfnamefont {V.}~\bibnamefont {Fatemi}},
  \bibinfo {author} {\bibfnamefont {S.}~\bibnamefont {Fang}}, \bibinfo {author}
  {\bibfnamefont {K.}~\bibnamefont {Watanabe}}, \bibinfo {author}
  {\bibfnamefont {T.}~\bibnamefont {Taniguchi}}, \bibinfo {author}
  {\bibfnamefont {E.}~\bibnamefont {Kaxiras}}, \ and\ \bibinfo {author}
  {\bibfnamefont {P.}~\bibnamefont {Jarillo-Herrero}},\ }\href@noop {}
  {\bibfield  {journal} {\bibinfo  {journal} {Nature}\ }\textbf {\bibinfo
  {volume} {556}},\ \bibinfo {pages} {43} (\bibinfo {year}
  {2018}{\natexlab{a}})}\BibitemShut {NoStop}%
\bibitem [{\citenamefont {Lieb}(1989)}]{lieb1989two}%
  \BibitemOpen
  \bibfield  {author} {\bibinfo {author} {\bibfnamefont {E.~H.}\ \bibnamefont
  {Lieb}},\ }\href@noop {} {\bibfield  {journal} {\bibinfo  {journal} {Physical
  review letters}\ }\textbf {\bibinfo {volume} {62}},\ \bibinfo {pages} {1201}
  (\bibinfo {year} {1989})}\BibitemShut {NoStop}%
\bibitem [{\citenamefont {Mielke}(1991)}]{mielke1991ferromagnetism}%
  \BibitemOpen
  \bibfield  {author} {\bibinfo {author} {\bibfnamefont {A.}~\bibnamefont
  {Mielke}},\ }\href@noop {} {\bibfield  {journal} {\bibinfo  {journal}
  {Journal of Physics A: Mathematical and General}\ }\textbf {\bibinfo {volume}
  {24}},\ \bibinfo {pages} {3311} (\bibinfo {year} {1991})}\BibitemShut
  {NoStop}%
\bibitem [{\citenamefont {Tanaka}\ and\ \citenamefont
  {Ueda}(2003)}]{tanaka2003stability}%
  \BibitemOpen
  \bibfield  {author} {\bibinfo {author} {\bibfnamefont {A.}~\bibnamefont
  {Tanaka}}\ and\ \bibinfo {author} {\bibfnamefont {H.}~\bibnamefont {Ueda}},\
  }\href@noop {} {\bibfield  {journal} {\bibinfo  {journal} {Physical review
  letters}\ }\textbf {\bibinfo {volume} {90}},\ \bibinfo {pages} {067204}
  (\bibinfo {year} {2003})}\BibitemShut {NoStop}%
\bibitem [{\citenamefont {Katsura}\ \emph {et~al.}(2010)\citenamefont
  {Katsura}, \citenamefont {Maruyama}, \citenamefont {Tanaka},\ and\
  \citenamefont {Tasaki}}]{katsura2010ferromagnetism}%
  \BibitemOpen
  \bibfield  {author} {\bibinfo {author} {\bibfnamefont {H.}~\bibnamefont
  {Katsura}}, \bibinfo {author} {\bibfnamefont {I.}~\bibnamefont {Maruyama}},
  \bibinfo {author} {\bibfnamefont {A.}~\bibnamefont {Tanaka}}, \ and\ \bibinfo
  {author} {\bibfnamefont {H.}~\bibnamefont {Tasaki}},\ }\href@noop {}
  {\bibfield  {journal} {\bibinfo  {journal} {EPL (Europhysics Letters)}\
  }\textbf {\bibinfo {volume} {91}},\ \bibinfo {pages} {57007} (\bibinfo {year}
  {2010})}\BibitemShut {NoStop}%
\bibitem [{\citenamefont {Tang}\ and\ \citenamefont
  {Fu}(2014)}]{tang2014strain}%
  \BibitemOpen
  \bibfield  {author} {\bibinfo {author} {\bibfnamefont {E.}~\bibnamefont
  {Tang}}\ and\ \bibinfo {author} {\bibfnamefont {L.}~\bibnamefont {Fu}},\
  }\href@noop {} {\bibfield  {journal} {\bibinfo  {journal} {Nature Physics}\
  }\textbf {\bibinfo {volume} {10}},\ \bibinfo {pages} {964} (\bibinfo {year}
  {2014})}\BibitemShut {NoStop}%
\bibitem [{\citenamefont {Cao}\ \emph {et~al.}(2018{\natexlab{b}})\citenamefont
  {Cao}, \citenamefont {Fatemi}, \citenamefont {Demir}, \citenamefont {Fang},
  \citenamefont {Tomarken}, \citenamefont {Luo}, \citenamefont
  {Sanchez-Yamagishi}, \citenamefont {Watanabe}, \citenamefont {Taniguchi},
  \citenamefont {Kaxiras} \emph {et~al.}}]{Cao2018-2}%
  \BibitemOpen
  \bibfield  {author} {\bibinfo {author} {\bibfnamefont {Y.}~\bibnamefont
  {Cao}}, \bibinfo {author} {\bibfnamefont {V.}~\bibnamefont {Fatemi}},
  \bibinfo {author} {\bibfnamefont {A.}~\bibnamefont {Demir}}, \bibinfo
  {author} {\bibfnamefont {S.}~\bibnamefont {Fang}}, \bibinfo {author}
  {\bibfnamefont {S.~L.}\ \bibnamefont {Tomarken}}, \bibinfo {author}
  {\bibfnamefont {J.~Y.}\ \bibnamefont {Luo}}, \bibinfo {author} {\bibfnamefont
  {J.~D.}\ \bibnamefont {Sanchez-Yamagishi}}, \bibinfo {author} {\bibfnamefont
  {K.}~\bibnamefont {Watanabe}}, \bibinfo {author} {\bibfnamefont
  {T.}~\bibnamefont {Taniguchi}}, \bibinfo {author} {\bibfnamefont
  {E.}~\bibnamefont {Kaxiras}},  \emph {et~al.},\ }\href@noop {} {\bibfield
  {journal} {\bibinfo  {journal} {Nature}\ }\textbf {\bibinfo {volume} {556}},\
  \bibinfo {pages} {80} (\bibinfo {year} {2018}{\natexlab{b}})}\BibitemShut
  {NoStop}%
\bibitem [{\citenamefont {Fogel}\ \emph {et~al.}(2006)\citenamefont {Fogel},
  \citenamefont {Buchstab}, \citenamefont {Bomze}, \citenamefont {Yuzephovich},
  \citenamefont {Mikhailov}, \citenamefont {Sipatov}, \citenamefont
  {Pashitskii}, \citenamefont {Shekhter},\ and\ \citenamefont
  {Jonson}}]{fogel2006direct}%
  \BibitemOpen
  \bibfield  {author} {\bibinfo {author} {\bibfnamefont {N.~Y.}\ \bibnamefont
  {Fogel}}, \bibinfo {author} {\bibfnamefont {E.}~\bibnamefont {Buchstab}},
  \bibinfo {author} {\bibfnamefont {Y.~V.}\ \bibnamefont {Bomze}}, \bibinfo
  {author} {\bibfnamefont {O.}~\bibnamefont {Yuzephovich}}, \bibinfo {author}
  {\bibfnamefont {M.~Y.}\ \bibnamefont {Mikhailov}}, \bibinfo {author}
  {\bibfnamefont {A.~Y.}\ \bibnamefont {Sipatov}}, \bibinfo {author}
  {\bibfnamefont {E.}~\bibnamefont {Pashitskii}}, \bibinfo {author}
  {\bibfnamefont {R.}~\bibnamefont {Shekhter}}, \ and\ \bibinfo {author}
  {\bibfnamefont {M.}~\bibnamefont {Jonson}},\ }\href@noop {} {\bibfield
  {journal} {\bibinfo  {journal} {Physical Review B}\ }\textbf {\bibinfo
  {volume} {73}},\ \bibinfo {pages} {161306} (\bibinfo {year}
  {2006})}\BibitemShut {NoStop}%
\bibitem [{\citenamefont {Mironov}\ \emph {et~al.}(1988)\citenamefont
  {Mironov}, \citenamefont {Savitskii}, \citenamefont {Sipatov}, \citenamefont
  {Fedorenko}, \citenamefont {Chirkin}, \citenamefont {Chistyakov},\ and\
  \citenamefont {Shpakovskaya}}]{mironov1988superconductivity}%
  \BibitemOpen
  \bibfield  {author} {\bibinfo {author} {\bibfnamefont {O.}~\bibnamefont
  {Mironov}}, \bibinfo {author} {\bibfnamefont {B.}~\bibnamefont {Savitskii}},
  \bibinfo {author} {\bibfnamefont {A.~Y.}\ \bibnamefont {Sipatov}}, \bibinfo
  {author} {\bibfnamefont {A.}~\bibnamefont {Fedorenko}}, \bibinfo {author}
  {\bibfnamefont {A.}~\bibnamefont {Chirkin}}, \bibinfo {author} {\bibfnamefont
  {S.}~\bibnamefont {Chistyakov}}, \ and\ \bibinfo {author} {\bibfnamefont
  {L.}~\bibnamefont {Shpakovskaya}},\ }\href@noop {} {\bibfield  {journal}
  {\bibinfo  {journal} {JETP Lett}\ }\textbf {\bibinfo {volume} {48}} (\bibinfo
  {year} {1988})}\BibitemShut {NoStop}%
\bibitem [{\citenamefont {Cao}\ \emph {et~al.}(2018{\natexlab{c}})\citenamefont
  {Cao}, \citenamefont {Fatemi}, \citenamefont {Fang}, \citenamefont
  {Watanabe}, \citenamefont {Taniguchi}, \citenamefont {Kaxiras},\ and\
  \citenamefont {Jarillo-Herrero}}]{Cao2018}%
  \BibitemOpen
  \bibfield  {author} {\bibinfo {author} {\bibfnamefont {Y.}~\bibnamefont
  {Cao}}, \bibinfo {author} {\bibfnamefont {V.}~\bibnamefont {Fatemi}},
  \bibinfo {author} {\bibfnamefont {S.}~\bibnamefont {Fang}}, \bibinfo {author}
  {\bibfnamefont {K.}~\bibnamefont {Watanabe}}, \bibinfo {author}
  {\bibfnamefont {T.}~\bibnamefont {Taniguchi}}, \bibinfo {author}
  {\bibfnamefont {E.}~\bibnamefont {Kaxiras}}, \ and\ \bibinfo {author}
  {\bibfnamefont {P.}~\bibnamefont {Jarillo-Herrero}},\ }\href
  {https://doi.org/10.1038/nature26160} {\bibfield  {journal} {\bibinfo
  {journal} {Nature}\ }\textbf {\bibinfo {volume} {556}},\ \bibinfo {pages} {43
  EP } (\bibinfo {year} {2018}{\natexlab{c}})},\ \bibinfo {note}
  {article}\BibitemShut {NoStop}%
\bibitem [{\citenamefont {Gong}\ \emph {et~al.}(2017)\citenamefont {Gong},
  \citenamefont {Li}, \citenamefont {Li}, \citenamefont {Ji}, \citenamefont
  {Stern}, \citenamefont {Xia}, \citenamefont {Cao}, \citenamefont {Bao},
  \citenamefont {Wang}, \citenamefont {Wang} \emph {et~al.}}]{Gong.nat2017}%
  \BibitemOpen
  \bibfield  {author} {\bibinfo {author} {\bibfnamefont {C.}~\bibnamefont
  {Gong}}, \bibinfo {author} {\bibfnamefont {L.}~\bibnamefont {Li}}, \bibinfo
  {author} {\bibfnamefont {Z.}~\bibnamefont {Li}}, \bibinfo {author}
  {\bibfnamefont {H.}~\bibnamefont {Ji}}, \bibinfo {author} {\bibfnamefont
  {A.}~\bibnamefont {Stern}}, \bibinfo {author} {\bibfnamefont
  {Y.}~\bibnamefont {Xia}}, \bibinfo {author} {\bibfnamefont {T.}~\bibnamefont
  {Cao}}, \bibinfo {author} {\bibfnamefont {W.}~\bibnamefont {Bao}}, \bibinfo
  {author} {\bibfnamefont {C.}~\bibnamefont {Wang}}, \bibinfo {author}
  {\bibfnamefont {Y.}~\bibnamefont {Wang}},  \emph {et~al.},\ }\href@noop {}
  {\bibfield  {journal} {\bibinfo  {journal} {Nature}\ }\textbf {\bibinfo
  {volume} {546}},\ \bibinfo {pages} {265} (\bibinfo {year}
  {2017})}\BibitemShut {NoStop}%
\bibitem [{\citenamefont {Huang}\ \emph {et~al.}(2017)\citenamefont {Huang},
  \citenamefont {Clark}, \citenamefont {Navarro-Moratalla}, \citenamefont
  {Klein}, \citenamefont {Cheng}, \citenamefont {Seyler}, \citenamefont
  {Zhong}, \citenamefont {Schmidgall}, \citenamefont {McGuire}, \citenamefont
  {Cobden} \emph {et~al.}}]{Huang.nat2017}%
  \BibitemOpen
  \bibfield  {author} {\bibinfo {author} {\bibfnamefont {B.}~\bibnamefont
  {Huang}}, \bibinfo {author} {\bibfnamefont {G.}~\bibnamefont {Clark}},
  \bibinfo {author} {\bibfnamefont {E.}~\bibnamefont {Navarro-Moratalla}},
  \bibinfo {author} {\bibfnamefont {D.~R.}\ \bibnamefont {Klein}}, \bibinfo
  {author} {\bibfnamefont {R.}~\bibnamefont {Cheng}}, \bibinfo {author}
  {\bibfnamefont {K.~L.}\ \bibnamefont {Seyler}}, \bibinfo {author}
  {\bibfnamefont {D.}~\bibnamefont {Zhong}}, \bibinfo {author} {\bibfnamefont
  {E.}~\bibnamefont {Schmidgall}}, \bibinfo {author} {\bibfnamefont {M.~A.}\
  \bibnamefont {McGuire}}, \bibinfo {author} {\bibfnamefont {D.~H.}\
  \bibnamefont {Cobden}},  \emph {et~al.},\ }\href@noop {} {\bibfield
  {journal} {\bibinfo  {journal} {Nature}\ }\textbf {\bibinfo {volume} {546}},\
  \bibinfo {pages} {270} (\bibinfo {year} {2017})}\BibitemShut {NoStop}%
\bibitem [{\citenamefont {Otrokov}\ \emph
  {et~al.}(2017{\natexlab{a}})\citenamefont {Otrokov}, \citenamefont
  {Menshchikova}, \citenamefont {Vergniory}, \citenamefont {Rusinov},
  \citenamefont {Vyazovskaya}, \citenamefont {Koroteev}, \citenamefont
  {Bihlmayer}, \citenamefont {Ernst}, \citenamefont {Echenique}, \citenamefont
  {Arnau} \emph {et~al.}}]{otrokov2017highly}%
  \BibitemOpen
  \bibfield  {author} {\bibinfo {author} {\bibfnamefont {M.}~\bibnamefont
  {Otrokov}}, \bibinfo {author} {\bibfnamefont {T.}~\bibnamefont
  {Menshchikova}}, \bibinfo {author} {\bibfnamefont {M.}~\bibnamefont
  {Vergniory}}, \bibinfo {author} {\bibfnamefont {I.}~\bibnamefont {Rusinov}},
  \bibinfo {author} {\bibfnamefont {A.~Y.}\ \bibnamefont {Vyazovskaya}},
  \bibinfo {author} {\bibfnamefont {Y.~M.}\ \bibnamefont {Koroteev}}, \bibinfo
  {author} {\bibfnamefont {G.}~\bibnamefont {Bihlmayer}}, \bibinfo {author}
  {\bibfnamefont {A.}~\bibnamefont {Ernst}}, \bibinfo {author} {\bibfnamefont
  {P.}~\bibnamefont {Echenique}}, \bibinfo {author} {\bibfnamefont
  {A.}~\bibnamefont {Arnau}},  \emph {et~al.},\ }\href@noop {} {\bibfield
  {journal} {\bibinfo  {journal} {2D Materials}\ }\textbf {\bibinfo {volume}
  {4}},\ \bibinfo {pages} {025082} (\bibinfo {year}
  {2017}{\natexlab{a}})}\BibitemShut {NoStop}%
\bibitem [{\citenamefont {Otrokov}\ \emph
  {et~al.}(2017{\natexlab{b}})\citenamefont {Otrokov}, \citenamefont
  {Menshchikova}, \citenamefont {Rusinov}, \citenamefont {Vergniory},
  \citenamefont {Kuznetsov},\ and\ \citenamefont
  {Chulkov}}]{Otrokov.jetpl2017}%
  \BibitemOpen
  \bibfield  {author} {\bibinfo {author} {\bibfnamefont {M.~M.}\ \bibnamefont
  {Otrokov}}, \bibinfo {author} {\bibfnamefont {T.~V.}\ \bibnamefont
  {Menshchikova}}, \bibinfo {author} {\bibfnamefont {I.~P.}\ \bibnamefont
  {Rusinov}}, \bibinfo {author} {\bibfnamefont {M.~G.}\ \bibnamefont
  {Vergniory}}, \bibinfo {author} {\bibfnamefont {V.~M.}\ \bibnamefont
  {Kuznetsov}}, \ and\ \bibinfo {author} {\bibfnamefont {E.~V.}\ \bibnamefont
  {Chulkov}},\ }\href@noop {} {\bibfield  {journal} {\bibinfo  {journal} {JETP
  Lett.}\ }\textbf {\bibinfo {volume} {105}},\ \bibinfo {pages} {297} (\bibinfo
  {year} {2017}{\natexlab{b}})}\BibitemShut {NoStop}%
\bibitem [{\citenamefont {Hirahara}\ \emph {et~al.}(2017)\citenamefont
  {Hirahara}, \citenamefont {Eremeev}, \citenamefont {Shirasawa}, \citenamefont
  {Okuyama}, \citenamefont {Kubo}, \citenamefont {Nakanishi}, \citenamefont
  {Akiyama}, \citenamefont {Takayama}, \citenamefont {Hajiri}, \citenamefont
  {Ideta} \emph {et~al.}}]{hirahara2017large}%
  \BibitemOpen
  \bibfield  {author} {\bibinfo {author} {\bibfnamefont {T.}~\bibnamefont
  {Hirahara}}, \bibinfo {author} {\bibfnamefont {S.~V.}\ \bibnamefont
  {Eremeev}}, \bibinfo {author} {\bibfnamefont {T.}~\bibnamefont {Shirasawa}},
  \bibinfo {author} {\bibfnamefont {Y.}~\bibnamefont {Okuyama}}, \bibinfo
  {author} {\bibfnamefont {T.}~\bibnamefont {Kubo}}, \bibinfo {author}
  {\bibfnamefont {R.}~\bibnamefont {Nakanishi}}, \bibinfo {author}
  {\bibfnamefont {R.}~\bibnamefont {Akiyama}}, \bibinfo {author} {\bibfnamefont
  {A.}~\bibnamefont {Takayama}}, \bibinfo {author} {\bibfnamefont
  {T.}~\bibnamefont {Hajiri}}, \bibinfo {author} {\bibfnamefont {S.-i.}\
  \bibnamefont {Ideta}},  \emph {et~al.},\ }\href@noop {} {\bibfield  {journal}
  {\bibinfo  {journal} {Nano letters}\ }\textbf {\bibinfo {volume} {17}},\
  \bibinfo {pages} {3493} (\bibinfo {year} {2017})}\BibitemShut {NoStop}%
\bibitem [{\citenamefont {Klein}\ \emph {et~al.}(2018)\citenamefont {Klein},
  \citenamefont {MacNeill}, \citenamefont {Lado}, \citenamefont {Soriano},
  \citenamefont {Navarro-Moratalla}, \citenamefont {Watanabe}, \citenamefont
  {Taniguchi}, \citenamefont {Manni}, \citenamefont {Canfield}, \citenamefont
  {Fern{\'a}ndez-Rossier} \emph {et~al.}}]{klein2018probing}%
  \BibitemOpen
  \bibfield  {author} {\bibinfo {author} {\bibfnamefont {D.~R.}\ \bibnamefont
  {Klein}}, \bibinfo {author} {\bibfnamefont {D.}~\bibnamefont {MacNeill}},
  \bibinfo {author} {\bibfnamefont {J.~L.}\ \bibnamefont {Lado}}, \bibinfo
  {author} {\bibfnamefont {D.}~\bibnamefont {Soriano}}, \bibinfo {author}
  {\bibfnamefont {E.}~\bibnamefont {Navarro-Moratalla}}, \bibinfo {author}
  {\bibfnamefont {K.}~\bibnamefont {Watanabe}}, \bibinfo {author}
  {\bibfnamefont {T.}~\bibnamefont {Taniguchi}}, \bibinfo {author}
  {\bibfnamefont {S.}~\bibnamefont {Manni}}, \bibinfo {author} {\bibfnamefont
  {P.}~\bibnamefont {Canfield}}, \bibinfo {author} {\bibfnamefont
  {J.}~\bibnamefont {Fern{\'a}ndez-Rossier}},  \emph {et~al.},\ }\href@noop {}
  {\bibfield  {journal} {\bibinfo  {journal} {Science}\ }\textbf {\bibinfo
  {volume} {360}},\ \bibinfo {pages} {1218} (\bibinfo {year}
  {2018})}\BibitemShut {NoStop}%
\bibitem [{\citenamefont {Song}\ \emph {et~al.}(2018)\citenamefont {Song},
  \citenamefont {Cai}, \citenamefont {Tu}, \citenamefont {Zhang}, \citenamefont
  {Huang}, \citenamefont {Wilson}, \citenamefont {Seyler}, \citenamefont {Zhu},
  \citenamefont {Taniguchi}, \citenamefont {Watanabe} \emph
  {et~al.}}]{song2018giant}%
  \BibitemOpen
  \bibfield  {author} {\bibinfo {author} {\bibfnamefont {T.}~\bibnamefont
  {Song}}, \bibinfo {author} {\bibfnamefont {X.}~\bibnamefont {Cai}}, \bibinfo
  {author} {\bibfnamefont {M.~W.-Y.}\ \bibnamefont {Tu}}, \bibinfo {author}
  {\bibfnamefont {X.}~\bibnamefont {Zhang}}, \bibinfo {author} {\bibfnamefont
  {B.}~\bibnamefont {Huang}}, \bibinfo {author} {\bibfnamefont {N.~P.}\
  \bibnamefont {Wilson}}, \bibinfo {author} {\bibfnamefont {K.~L.}\
  \bibnamefont {Seyler}}, \bibinfo {author} {\bibfnamefont {L.}~\bibnamefont
  {Zhu}}, \bibinfo {author} {\bibfnamefont {T.}~\bibnamefont {Taniguchi}},
  \bibinfo {author} {\bibfnamefont {K.}~\bibnamefont {Watanabe}},  \emph
  {et~al.},\ }\href@noop {} {\bibfield  {journal} {\bibinfo  {journal}
  {Science}\ }\textbf {\bibinfo {volume} {360}},\ \bibinfo {pages} {1214}
  (\bibinfo {year} {2018})}\BibitemShut {NoStop}%
\bibitem [{\citenamefont {Wang}\ \emph {et~al.}(2018)\citenamefont {Wang},
  \citenamefont {Guti{\'e}rrez-Lezama}, \citenamefont {Ubrig}, \citenamefont
  {Kroner}, \citenamefont {Gibertini}, \citenamefont {Taniguchi}, \citenamefont
  {Watanabe}, \citenamefont {Imamo{\u{g}}lu}, \citenamefont {Giannini},\ and\
  \citenamefont {Morpurgo}}]{wang2018very}%
  \BibitemOpen
  \bibfield  {author} {\bibinfo {author} {\bibfnamefont {Z.}~\bibnamefont
  {Wang}}, \bibinfo {author} {\bibfnamefont {I.}~\bibnamefont
  {Guti{\'e}rrez-Lezama}}, \bibinfo {author} {\bibfnamefont {N.}~\bibnamefont
  {Ubrig}}, \bibinfo {author} {\bibfnamefont {M.}~\bibnamefont {Kroner}},
  \bibinfo {author} {\bibfnamefont {M.}~\bibnamefont {Gibertini}}, \bibinfo
  {author} {\bibfnamefont {T.}~\bibnamefont {Taniguchi}}, \bibinfo {author}
  {\bibfnamefont {K.}~\bibnamefont {Watanabe}}, \bibinfo {author}
  {\bibfnamefont {A.}~\bibnamefont {Imamo{\u{g}}lu}}, \bibinfo {author}
  {\bibfnamefont {E.}~\bibnamefont {Giannini}}, \ and\ \bibinfo {author}
  {\bibfnamefont {A.~F.}\ \bibnamefont {Morpurgo}},\ }\href@noop {} {\bibfield
  {journal} {\bibinfo  {journal} {Nature communications}\ }\textbf {\bibinfo
  {volume} {9}},\ \bibinfo {pages} {2516} (\bibinfo {year} {2018})}\BibitemShut
  {NoStop}%
\bibitem [{\citenamefont {Mong}\ \emph {et~al.}(2010)\citenamefont {Mong},
  \citenamefont {Essin},\ and\ \citenamefont
  {Moore}}]{mong2010antiferromagnetic}%
  \BibitemOpen
  \bibfield  {author} {\bibinfo {author} {\bibfnamefont {R.~S.}\ \bibnamefont
  {Mong}}, \bibinfo {author} {\bibfnamefont {A.~M.}\ \bibnamefont {Essin}}, \
  and\ \bibinfo {author} {\bibfnamefont {J.~E.}\ \bibnamefont {Moore}},\
  }\href@noop {} {\bibfield  {journal} {\bibinfo  {journal} {Physical Review
  B}\ }\textbf {\bibinfo {volume} {81}},\ \bibinfo {pages} {245209} (\bibinfo
  {year} {2010})}\BibitemShut {NoStop}%
\bibitem [{\citenamefont {McGuire}\ \emph {et~al.}(2015)\citenamefont
  {McGuire}, \citenamefont {Dixit}, \citenamefont {Cooper},\ and\ \citenamefont
  {Sales}}]{mcguire2015coupling}%
  \BibitemOpen
  \bibfield  {author} {\bibinfo {author} {\bibfnamefont {M.~A.}\ \bibnamefont
  {McGuire}}, \bibinfo {author} {\bibfnamefont {H.}~\bibnamefont {Dixit}},
  \bibinfo {author} {\bibfnamefont {V.~R.}\ \bibnamefont {Cooper}}, \ and\
  \bibinfo {author} {\bibfnamefont {B.~C.}\ \bibnamefont {Sales}},\ }\href@noop
  {} {\bibfield  {journal} {\bibinfo  {journal} {Chemistry of Materials}\
  }\textbf {\bibinfo {volume} {27}},\ \bibinfo {pages} {612} (\bibinfo {year}
  {2015})}\BibitemShut {NoStop}%
\bibitem [{\citenamefont {Sass}\ \emph {et~al.}(2019)\citenamefont {Sass},
  \citenamefont {Ge}, \citenamefont {Yan}, \citenamefont {Obeysekera},
  \citenamefont {Yang},\ and\ \citenamefont {Wu}}]{sass2019magnetic}%
  \BibitemOpen
  \bibfield  {author} {\bibinfo {author} {\bibfnamefont {P.~M.}\ \bibnamefont
  {Sass}}, \bibinfo {author} {\bibfnamefont {W.}~\bibnamefont {Ge}}, \bibinfo
  {author} {\bibfnamefont {J.}~\bibnamefont {Yan}}, \bibinfo {author}
  {\bibfnamefont {D.}~\bibnamefont {Obeysekera}}, \bibinfo {author}
  {\bibfnamefont {J.}~\bibnamefont {Yang}}, \ and\ \bibinfo {author}
  {\bibfnamefont {W.}~\bibnamefont {Wu}},\ }\href@noop {} {\bibfield  {journal}
  {\bibinfo  {journal} {arXiv preprint arXiv:1910.06488}\ } (\bibinfo {year}
  {2019})}\BibitemShut {NoStop}%
\bibitem [{\citenamefont {Yasuda}\ \emph {et~al.}(2017)\citenamefont {Yasuda},
  \citenamefont {Mogi}, \citenamefont {Yoshimi}, \citenamefont {Tsukazaki},
  \citenamefont {Takahashi}, \citenamefont {Kawasaki}, \citenamefont {Kagawa},\
  and\ \citenamefont {Tokura}}]{yasuda2017quantized}%
  \BibitemOpen
  \bibfield  {author} {\bibinfo {author} {\bibfnamefont {K.}~\bibnamefont
  {Yasuda}}, \bibinfo {author} {\bibfnamefont {M.}~\bibnamefont {Mogi}},
  \bibinfo {author} {\bibfnamefont {R.}~\bibnamefont {Yoshimi}}, \bibinfo
  {author} {\bibfnamefont {A.}~\bibnamefont {Tsukazaki}}, \bibinfo {author}
  {\bibfnamefont {K.}~\bibnamefont {Takahashi}}, \bibinfo {author}
  {\bibfnamefont {M.}~\bibnamefont {Kawasaki}}, \bibinfo {author}
  {\bibfnamefont {F.}~\bibnamefont {Kagawa}}, \ and\ \bibinfo {author}
  {\bibfnamefont {Y.}~\bibnamefont {Tokura}},\ }\href@noop {} {\bibfield
  {journal} {\bibinfo  {journal} {Science}\ }\textbf {\bibinfo {volume}
  {358}},\ \bibinfo {pages} {1311} (\bibinfo {year} {2017})}\BibitemShut
  {NoStop}%
\bibitem [{\citenamefont {Rosen}\ \emph {et~al.}(2017)\citenamefont {Rosen},
  \citenamefont {Fox}, \citenamefont {Kou}, \citenamefont {Pan}, \citenamefont
  {Wang},\ and\ \citenamefont {Goldhaber-Gordon}}]{rosen2017chiral}%
  \BibitemOpen
  \bibfield  {author} {\bibinfo {author} {\bibfnamefont {I.~T.}\ \bibnamefont
  {Rosen}}, \bibinfo {author} {\bibfnamefont {E.~J.}\ \bibnamefont {Fox}},
  \bibinfo {author} {\bibfnamefont {X.}~\bibnamefont {Kou}}, \bibinfo {author}
  {\bibfnamefont {L.}~\bibnamefont {Pan}}, \bibinfo {author} {\bibfnamefont
  {K.~L.}\ \bibnamefont {Wang}}, \ and\ \bibinfo {author} {\bibfnamefont
  {D.}~\bibnamefont {Goldhaber-Gordon}},\ }\href@noop {} {\bibfield  {journal}
  {\bibinfo  {journal} {npj Quantum Materials}\ }\textbf {\bibinfo {volume}
  {2}},\ \bibinfo {pages} {69} (\bibinfo {year} {2017})}\BibitemShut {NoStop}%
\bibitem [{\citenamefont {Eremeev}\ \emph {et~al.}(2012)\citenamefont
  {Eremeev}, \citenamefont {Landolt}, \citenamefont {Menshchikova},
  \citenamefont {Slomski}, \citenamefont {Koroteev}, \citenamefont {Aliev},
  \citenamefont {Babanly}, \citenamefont {Henk}, \citenamefont {Ernst},
  \citenamefont {Patthey} \emph {et~al.}}]{eremeev2012atom}%
  \BibitemOpen
  \bibfield  {author} {\bibinfo {author} {\bibfnamefont {S.~V.}\ \bibnamefont
  {Eremeev}}, \bibinfo {author} {\bibfnamefont {G.}~\bibnamefont {Landolt}},
  \bibinfo {author} {\bibfnamefont {T.~V.}\ \bibnamefont {Menshchikova}},
  \bibinfo {author} {\bibfnamefont {B.}~\bibnamefont {Slomski}}, \bibinfo
  {author} {\bibfnamefont {Y.~M.}\ \bibnamefont {Koroteev}}, \bibinfo {author}
  {\bibfnamefont {Z.~S.}\ \bibnamefont {Aliev}}, \bibinfo {author}
  {\bibfnamefont {M.~B.}\ \bibnamefont {Babanly}}, \bibinfo {author}
  {\bibfnamefont {J.}~\bibnamefont {Henk}}, \bibinfo {author} {\bibfnamefont
  {A.}~\bibnamefont {Ernst}}, \bibinfo {author} {\bibfnamefont
  {L.}~\bibnamefont {Patthey}},  \emph {et~al.},\ }\href@noop {} {\bibfield
  {journal} {\bibinfo  {journal} {Nature communications}\ }\textbf {\bibinfo
  {volume} {3}},\ \bibinfo {pages} {635} (\bibinfo {year} {2012})}\BibitemShut
  {NoStop}%
\bibitem [{\citenamefont {Otrokov}\ \emph
  {et~al.}(2019{\natexlab{a}})\citenamefont {Otrokov}, \citenamefont
  {Klimovskikh}, \citenamefont {Bentmann}, \citenamefont {Estyunin},
  \citenamefont {Zeugner}, \citenamefont {Aliev}, \citenamefont {Ga{\ss}},
  \citenamefont {Wolter}, \citenamefont {Koroleva}, \citenamefont {Shikin}
  \emph {et~al.}}]{otrokov2019prediction}%
  \BibitemOpen
  \bibfield  {author} {\bibinfo {author} {\bibfnamefont {M.}~\bibnamefont
  {Otrokov}}, \bibinfo {author} {\bibfnamefont {I.}~\bibnamefont
  {Klimovskikh}}, \bibinfo {author} {\bibfnamefont {H.}~\bibnamefont
  {Bentmann}}, \bibinfo {author} {\bibfnamefont {D.}~\bibnamefont {Estyunin}},
  \bibinfo {author} {\bibfnamefont {A.}~\bibnamefont {Zeugner}}, \bibinfo
  {author} {\bibfnamefont {Z.}~\bibnamefont {Aliev}}, \bibinfo {author}
  {\bibfnamefont {S.}~\bibnamefont {Ga{\ss}}}, \bibinfo {author} {\bibfnamefont
  {A.}~\bibnamefont {Wolter}}, \bibinfo {author} {\bibfnamefont
  {A.}~\bibnamefont {Koroleva}}, \bibinfo {author} {\bibfnamefont
  {A.}~\bibnamefont {Shikin}},  \emph {et~al.},\ }\href@noop {} {\bibfield
  {journal} {\bibinfo  {journal} {Nature}\ }\textbf {\bibinfo {volume} {576}},\
  \bibinfo {pages} {416} (\bibinfo {year} {2019}{\natexlab{a}})}\BibitemShut
  {NoStop}%
\bibitem [{\citenamefont {Otrokov}\ \emph
  {et~al.}(2019{\natexlab{b}})\citenamefont {Otrokov}, \citenamefont {Rusinov},
  \citenamefont {Blanco-Rey}, \citenamefont {Hoffmann}, \citenamefont
  {Vyazovskaya}, \citenamefont {Eremeev}, \citenamefont {Ernst}, \citenamefont
  {Echenique}, \citenamefont {Arnau},\ and\ \citenamefont
  {Chulkov}}]{Otrokov.prl2019}%
  \BibitemOpen
  \bibfield  {author} {\bibinfo {author} {\bibfnamefont {M.~M.}\ \bibnamefont
  {Otrokov}}, \bibinfo {author} {\bibfnamefont {I.~P.}\ \bibnamefont
  {Rusinov}}, \bibinfo {author} {\bibfnamefont {M.}~\bibnamefont {Blanco-Rey}},
  \bibinfo {author} {\bibfnamefont {M.}~\bibnamefont {Hoffmann}}, \bibinfo
  {author} {\bibfnamefont {A.~Y.}\ \bibnamefont {Vyazovskaya}}, \bibinfo
  {author} {\bibfnamefont {S.~V.}\ \bibnamefont {Eremeev}}, \bibinfo {author}
  {\bibfnamefont {A.}~\bibnamefont {Ernst}}, \bibinfo {author} {\bibfnamefont
  {P.~M.}\ \bibnamefont {Echenique}}, \bibinfo {author} {\bibfnamefont
  {A.}~\bibnamefont {Arnau}}, \ and\ \bibinfo {author} {\bibfnamefont {E.~V.}\
  \bibnamefont {Chulkov}},\ }\href {\doibase 10.1103/PhysRevLett.122.107202}
  {\bibfield  {journal} {\bibinfo  {journal} {Phys. Rev. Lett.}\ }\textbf
  {\bibinfo {volume} {122}},\ \bibinfo {pages} {107202} (\bibinfo {year}
  {2019}{\natexlab{b}})}\BibitemShut {NoStop}%
\bibitem [{\citenamefont {Klimovskikh}\ \emph {et~al.}(2019)\citenamefont
  {Klimovskikh}, \citenamefont {Otrokov}, \citenamefont {Estyunin},
  \citenamefont {Eremeev}, \citenamefont {Filnov}, \citenamefont {Koroleva},
  \citenamefont {Shevchenko}, \citenamefont {Voroshnin}, \citenamefont
  {Rusinov}, \citenamefont {Blanco-Rey} \emph
  {et~al.}}]{klimovskikh2019variety}%
  \BibitemOpen
  \bibfield  {author} {\bibinfo {author} {\bibfnamefont {I.}~\bibnamefont
  {Klimovskikh}}, \bibinfo {author} {\bibfnamefont {M.}~\bibnamefont
  {Otrokov}}, \bibinfo {author} {\bibfnamefont {D.}~\bibnamefont {Estyunin}},
  \bibinfo {author} {\bibfnamefont {S.}~\bibnamefont {Eremeev}}, \bibinfo
  {author} {\bibfnamefont {S.}~\bibnamefont {Filnov}}, \bibinfo {author}
  {\bibfnamefont {A.}~\bibnamefont {Koroleva}}, \bibinfo {author}
  {\bibfnamefont {E.}~\bibnamefont {Shevchenko}}, \bibinfo {author}
  {\bibfnamefont {V.}~\bibnamefont {Voroshnin}}, \bibinfo {author}
  {\bibfnamefont {I.}~\bibnamefont {Rusinov}}, \bibinfo {author} {\bibfnamefont
  {M.}~\bibnamefont {Blanco-Rey}},  \emph {et~al.},\ }\href@noop {} {\bibfield
  {journal} {\bibinfo  {journal} {arXiv preprint arXiv:1910.11653}\ } (\bibinfo
  {year} {2019})}\BibitemShut {NoStop}%
\bibitem [{\citenamefont {Wu}\ \emph {et~al.}(2019)\citenamefont {Wu},
  \citenamefont {Liu}, \citenamefont {Sasase}, \citenamefont {Ienaga},
  \citenamefont {Obata}, \citenamefont {Yukawa}, \citenamefont {Horiba},
  \citenamefont {Kumigashira}, \citenamefont {Okuma}, \citenamefont {Inoshita}
  \emph {et~al.}}]{wu2019natural}%
  \BibitemOpen
  \bibfield  {author} {\bibinfo {author} {\bibfnamefont {J.}~\bibnamefont
  {Wu}}, \bibinfo {author} {\bibfnamefont {F.}~\bibnamefont {Liu}}, \bibinfo
  {author} {\bibfnamefont {M.}~\bibnamefont {Sasase}}, \bibinfo {author}
  {\bibfnamefont {K.}~\bibnamefont {Ienaga}}, \bibinfo {author} {\bibfnamefont
  {Y.}~\bibnamefont {Obata}}, \bibinfo {author} {\bibfnamefont
  {R.}~\bibnamefont {Yukawa}}, \bibinfo {author} {\bibfnamefont
  {K.}~\bibnamefont {Horiba}}, \bibinfo {author} {\bibfnamefont
  {H.}~\bibnamefont {Kumigashira}}, \bibinfo {author} {\bibfnamefont
  {S.}~\bibnamefont {Okuma}}, \bibinfo {author} {\bibfnamefont
  {T.}~\bibnamefont {Inoshita}},  \emph {et~al.},\ }\href@noop {} {\bibfield
  {journal} {\bibinfo  {journal} {Science advances}\ }\textbf {\bibinfo
  {volume} {5}},\ \bibinfo {pages} {eaax9989} (\bibinfo {year}
  {2019})}\BibitemShut {NoStop}%
\bibitem [{\citenamefont {Chen}\ \emph {et~al.}(2019)\citenamefont {Chen},
  \citenamefont {Fei}, \citenamefont {Zhang}, \citenamefont {Zhang},
  \citenamefont {Liu}, \citenamefont {Zhang}, \citenamefont {Wang},
  \citenamefont {Wei}, \citenamefont {Zhang}, \citenamefont {Zuo} \emph
  {et~al.}}]{chen2019intrinsic}%
  \BibitemOpen
  \bibfield  {author} {\bibinfo {author} {\bibfnamefont {B.}~\bibnamefont
  {Chen}}, \bibinfo {author} {\bibfnamefont {F.}~\bibnamefont {Fei}}, \bibinfo
  {author} {\bibfnamefont {D.}~\bibnamefont {Zhang}}, \bibinfo {author}
  {\bibfnamefont {B.}~\bibnamefont {Zhang}}, \bibinfo {author} {\bibfnamefont
  {W.}~\bibnamefont {Liu}}, \bibinfo {author} {\bibfnamefont {S.}~\bibnamefont
  {Zhang}}, \bibinfo {author} {\bibfnamefont {P.}~\bibnamefont {Wang}},
  \bibinfo {author} {\bibfnamefont {B.}~\bibnamefont {Wei}}, \bibinfo {author}
  {\bibfnamefont {Y.}~\bibnamefont {Zhang}}, \bibinfo {author} {\bibfnamefont
  {Z.}~\bibnamefont {Zuo}},  \emph {et~al.},\ }\href@noop {} {\bibfield
  {journal} {\bibinfo  {journal} {Nature communications}\ }\textbf {\bibinfo
  {volume} {10}},\ \bibinfo {pages} {1} (\bibinfo {year} {2019})}\BibitemShut
  {NoStop}%
\bibitem [{\citenamefont {Eremeev}\ \emph {et~al.}(2017)\citenamefont
  {Eremeev}, \citenamefont {Otrokov},\ and\ \citenamefont
  {Chulkov}}]{eremeev2017competing}%
  \BibitemOpen
  \bibfield  {author} {\bibinfo {author} {\bibfnamefont {S.}~\bibnamefont
  {Eremeev}}, \bibinfo {author} {\bibfnamefont {M.}~\bibnamefont {Otrokov}}, \
  and\ \bibinfo {author} {\bibfnamefont {E.}~\bibnamefont {Chulkov}},\
  }\href@noop {} {\bibfield  {journal} {\bibinfo  {journal} {Journal of Alloys
  and Compounds}\ }\textbf {\bibinfo {volume} {709}},\ \bibinfo {pages} {172}
  (\bibinfo {year} {2017})}\BibitemShut {NoStop}%
\bibitem [{\citenamefont {Viana}\ and\ \citenamefont
  {de~Sousa}(2007)}]{viana2007anisotropy}%
  \BibitemOpen
  \bibfield  {author} {\bibinfo {author} {\bibfnamefont {J.~R.}\ \bibnamefont
  {Viana}}\ and\ \bibinfo {author} {\bibfnamefont {J.~R.}\ \bibnamefont
  {de~Sousa}},\ }\href@noop {} {\bibfield  {journal} {\bibinfo  {journal}
  {Physical Review B}\ }\textbf {\bibinfo {volume} {75}},\ \bibinfo {pages}
  {052403} (\bibinfo {year} {2007})}\BibitemShut {NoStop}%
\bibitem [{\citenamefont {Mermin}\ and\ \citenamefont
  {Wagner}(1966)}]{mermin1966absence}%
  \BibitemOpen
  \bibfield  {author} {\bibinfo {author} {\bibfnamefont {N.~D.}\ \bibnamefont
  {Mermin}}\ and\ \bibinfo {author} {\bibfnamefont {H.}~\bibnamefont
  {Wagner}},\ }\href@noop {} {\bibfield  {journal} {\bibinfo  {journal}
  {Physical Review Letters}\ }\textbf {\bibinfo {volume} {17}},\ \bibinfo
  {pages} {1133} (\bibinfo {year} {1966})}\BibitemShut {NoStop}%
\bibitem [{\citenamefont {Efetov}\ and\ \citenamefont
  {Larkin}(1975)}]{efetov1975pairing}%
  \BibitemOpen
  \bibfield  {author} {\bibinfo {author} {\bibfnamefont {K.}~\bibnamefont
  {Efetov}}\ and\ \bibinfo {author} {\bibfnamefont {A.}~\bibnamefont
  {Larkin}},\ }\href@noop {} {\bibfield  {journal} {\bibinfo  {journal} {Zh.
  Eksp. Teor. Fiz}\ }\textbf {\bibinfo {volume} {68}},\ \bibinfo {pages} {155}
  (\bibinfo {year} {1975})}\BibitemShut {NoStop}%
\bibitem [{\citenamefont {Bl{\"o}chl}(1994)}]{blochl1994projector}%
  \BibitemOpen
  \bibfield  {author} {\bibinfo {author} {\bibfnamefont {P.~E.}\ \bibnamefont
  {Bl{\"o}chl}},\ }\href@noop {} {\bibfield  {journal} {\bibinfo  {journal}
  {Physical review B}\ }\textbf {\bibinfo {volume} {50}},\ \bibinfo {pages}
  {17953} (\bibinfo {year} {1994})}\BibitemShut {NoStop}%
\bibitem [{\citenamefont {Kresse}\ and\ \citenamefont
  {Hafner}(1993)}]{kresse1993ab}%
  \BibitemOpen
  \bibfield  {author} {\bibinfo {author} {\bibfnamefont {G.}~\bibnamefont
  {Kresse}}\ and\ \bibinfo {author} {\bibfnamefont {J.}~\bibnamefont
  {Hafner}},\ }\href@noop {} {\bibfield  {journal} {\bibinfo  {journal}
  {Physical Review B}\ }\textbf {\bibinfo {volume} {47}},\ \bibinfo {pages}
  {558} (\bibinfo {year} {1993})}\BibitemShut {NoStop}%
\bibitem [{\citenamefont {Kresse}\ and\ \citenamefont
  {Furthm{\"u}ller}(1996)}]{kresse1996efficient}%
  \BibitemOpen
  \bibfield  {author} {\bibinfo {author} {\bibfnamefont {G.}~\bibnamefont
  {Kresse}}\ and\ \bibinfo {author} {\bibfnamefont {J.}~\bibnamefont
  {Furthm{\"u}ller}},\ }\href@noop {} {\bibfield  {journal} {\bibinfo
  {journal} {Physical review B}\ }\textbf {\bibinfo {volume} {54}},\ \bibinfo
  {pages} {11169} (\bibinfo {year} {1996})}\BibitemShut {NoStop}%
\bibitem [{\citenamefont {Kresse}\ \emph {et~al.}(1996)\citenamefont {Kresse}
  \emph {et~al.}}]{kresse1996g}%
  \BibitemOpen
  \bibfield  {author} {\bibinfo {author} {\bibfnamefont {G.}~\bibnamefont
  {Kresse}} \emph {et~al.},\ }\href@noop {} {\bibfield  {journal} {\bibinfo
  {journal} {Comput. Mater. Sci.}\ }\textbf {\bibinfo {volume} {6}},\ \bibinfo
  {pages} {15} (\bibinfo {year} {1996})}\BibitemShut {NoStop}%
\bibitem [{\citenamefont {Perdew}\ \emph {et~al.}(1996)\citenamefont {Perdew},
  \citenamefont {Burke},\ and\ \citenamefont
  {Ernzerhof}}]{perdew1996generalized}%
  \BibitemOpen
  \bibfield  {author} {\bibinfo {author} {\bibfnamefont {J.~P.}\ \bibnamefont
  {Perdew}}, \bibinfo {author} {\bibfnamefont {K.}~\bibnamefont {Burke}}, \
  and\ \bibinfo {author} {\bibfnamefont {M.}~\bibnamefont {Ernzerhof}},\
  }\href@noop {} {\bibfield  {journal} {\bibinfo  {journal} {Physical review
  letters}\ }\textbf {\bibinfo {volume} {77}},\ \bibinfo {pages} {3865}
  (\bibinfo {year} {1996})}\BibitemShut {NoStop}%
\bibitem [{\citenamefont {Koelling}\ and\ \citenamefont
  {Harmon}(1977)}]{koelling1977technique}%
  \BibitemOpen
  \bibfield  {author} {\bibinfo {author} {\bibfnamefont {D.}~\bibnamefont
  {Koelling}}\ and\ \bibinfo {author} {\bibfnamefont {B.}~\bibnamefont
  {Harmon}},\ }\href@noop {} {\bibfield  {journal} {\bibinfo  {journal}
  {Journal of Physics C: Solid State Physics}\ }\textbf {\bibinfo {volume}
  {10}},\ \bibinfo {pages} {3107} (\bibinfo {year} {1977})}\BibitemShut
  {NoStop}%
\bibitem [{\citenamefont {Grimme}\ \emph {et~al.}(2010)\citenamefont {Grimme},
  \citenamefont {Antony}, \citenamefont {Ehrlich},\ and\ \citenamefont
  {Krieg}}]{grimme2010consistent}%
  \BibitemOpen
  \bibfield  {author} {\bibinfo {author} {\bibfnamefont {S.}~\bibnamefont
  {Grimme}}, \bibinfo {author} {\bibfnamefont {J.}~\bibnamefont {Antony}},
  \bibinfo {author} {\bibfnamefont {S.}~\bibnamefont {Ehrlich}}, \ and\
  \bibinfo {author} {\bibfnamefont {H.}~\bibnamefont {Krieg}},\ }\href@noop {}
  {\bibfield  {journal} {\bibinfo  {journal} {The Journal of chemical physics}\
  }\textbf {\bibinfo {volume} {132}},\ \bibinfo {pages} {154104} (\bibinfo
  {year} {2010})}\BibitemShut {NoStop}%
\bibitem [{\citenamefont {Anisimov}\ \emph {et~al.}(1991)\citenamefont
  {Anisimov}, \citenamefont {Zaanen},\ and\ \citenamefont
  {Andersen}}]{anisimov1991band}%
  \BibitemOpen
  \bibfield  {author} {\bibinfo {author} {\bibfnamefont {V.~I.}\ \bibnamefont
  {Anisimov}}, \bibinfo {author} {\bibfnamefont {J.}~\bibnamefont {Zaanen}}, \
  and\ \bibinfo {author} {\bibfnamefont {O.~K.}\ \bibnamefont {Andersen}},\
  }\href@noop {} {\bibfield  {journal} {\bibinfo  {journal} {Physical Review
  B}\ }\textbf {\bibinfo {volume} {44}},\ \bibinfo {pages} {943} (\bibinfo
  {year} {1991})}\BibitemShut {NoStop}%
\bibitem [{\citenamefont {Dudarev}\ \emph {et~al.}(1998)\citenamefont
  {Dudarev}, \citenamefont {Botton}, \citenamefont {Savrasov}, \citenamefont
  {Humphreys},\ and\ \citenamefont {Sutton}}]{dudarev1998electron}%
  \BibitemOpen
  \bibfield  {author} {\bibinfo {author} {\bibfnamefont {S.}~\bibnamefont
  {Dudarev}}, \bibinfo {author} {\bibfnamefont {G.}~\bibnamefont {Botton}},
  \bibinfo {author} {\bibfnamefont {S.}~\bibnamefont {Savrasov}}, \bibinfo
  {author} {\bibfnamefont {C.}~\bibnamefont {Humphreys}}, \ and\ \bibinfo
  {author} {\bibfnamefont {A.}~\bibnamefont {Sutton}},\ }\href@noop {}
  {\bibfield  {journal} {\bibinfo  {journal} {Physical Review B}\ }\textbf
  {\bibinfo {volume} {57}},\ \bibinfo {pages} {1505} (\bibinfo {year}
  {1998})}\BibitemShut {NoStop}%
\bibitem [{\citenamefont {Cococcioni}\ and\ \citenamefont
  {De~Gironcoli}(2005)}]{cococcioni2005linear}%
  \BibitemOpen
  \bibfield  {author} {\bibinfo {author} {\bibfnamefont {M.}~\bibnamefont
  {Cococcioni}}\ and\ \bibinfo {author} {\bibfnamefont {S.}~\bibnamefont
  {De~Gironcoli}},\ }\href@noop {} {\bibfield  {journal} {\bibinfo  {journal}
  {Physical Review B}\ }\textbf {\bibinfo {volume} {71}},\ \bibinfo {pages}
  {035105} (\bibinfo {year} {2005})}\BibitemShut {NoStop}%
\bibitem [{\citenamefont {Soluyanov}\ and\ \citenamefont
  {Vanderbilt}(2011)}]{soluyanov2011computing}%
  \BibitemOpen
  \bibfield  {author} {\bibinfo {author} {\bibfnamefont {A.~A.}\ \bibnamefont
  {Soluyanov}}\ and\ \bibinfo {author} {\bibfnamefont {D.}~\bibnamefont
  {Vanderbilt}},\ }\href@noop {} {\bibfield  {journal} {\bibinfo  {journal}
  {Physical Review B}\ }\textbf {\bibinfo {volume} {83}},\ \bibinfo {pages}
  {235401} (\bibinfo {year} {2011})}\BibitemShut {NoStop}%
\bibitem [{\citenamefont {Gresch}\ \emph {et~al.}(2017)\citenamefont {Gresch},
  \citenamefont {Autes}, \citenamefont {Yazyev}, \citenamefont {Troyer},
  \citenamefont {Vanderbilt}, \citenamefont {Bernevig},\ and\ \citenamefont
  {Soluyanov}}]{gresch2017z2pack}%
  \BibitemOpen
  \bibfield  {author} {\bibinfo {author} {\bibfnamefont {D.}~\bibnamefont
  {Gresch}}, \bibinfo {author} {\bibfnamefont {G.}~\bibnamefont {Autes}},
  \bibinfo {author} {\bibfnamefont {O.~V.}\ \bibnamefont {Yazyev}}, \bibinfo
  {author} {\bibfnamefont {M.}~\bibnamefont {Troyer}}, \bibinfo {author}
  {\bibfnamefont {D.}~\bibnamefont {Vanderbilt}}, \bibinfo {author}
  {\bibfnamefont {B.~A.}\ \bibnamefont {Bernevig}}, \ and\ \bibinfo {author}
  {\bibfnamefont {A.~A.}\ \bibnamefont {Soluyanov}},\ }\href@noop {} {\bibfield
   {journal} {\bibinfo  {journal} {Physical Review B}\ }\textbf {\bibinfo
  {volume} {95}},\ \bibinfo {pages} {075146} (\bibinfo {year}
  {2017})}\BibitemShut {NoStop}%
\bibitem [{\citenamefont {Marzari}\ and\ \citenamefont
  {Vanderbilt}(1997)}]{marzari1997maximally}%
  \BibitemOpen
  \bibfield  {author} {\bibinfo {author} {\bibfnamefont {N.}~\bibnamefont
  {Marzari}}\ and\ \bibinfo {author} {\bibfnamefont {D.}~\bibnamefont
  {Vanderbilt}},\ }\href@noop {} {\bibfield  {journal} {\bibinfo  {journal}
  {Physical review B}\ }\textbf {\bibinfo {volume} {56}},\ \bibinfo {pages}
  {12847} (\bibinfo {year} {1997})}\BibitemShut {NoStop}%
\bibitem [{\citenamefont {Mostofi}\ \emph {et~al.}(2008)\citenamefont
  {Mostofi}, \citenamefont {Yates}, \citenamefont {Lee}, \citenamefont {Souza},
  \citenamefont {Vanderbilt},\ and\ \citenamefont
  {Marzari}}]{mostofi2008wannier90}%
  \BibitemOpen
  \bibfield  {author} {\bibinfo {author} {\bibfnamefont {A.~A.}\ \bibnamefont
  {Mostofi}}, \bibinfo {author} {\bibfnamefont {J.~R.}\ \bibnamefont {Yates}},
  \bibinfo {author} {\bibfnamefont {Y.-S.}\ \bibnamefont {Lee}}, \bibinfo
  {author} {\bibfnamefont {I.}~\bibnamefont {Souza}}, \bibinfo {author}
  {\bibfnamefont {D.}~\bibnamefont {Vanderbilt}}, \ and\ \bibinfo {author}
  {\bibfnamefont {N.}~\bibnamefont {Marzari}},\ }\href@noop {} {\bibfield
  {journal} {\bibinfo  {journal} {Computer physics communications}\ }\textbf
  {\bibinfo {volume} {178}},\ \bibinfo {pages} {685} (\bibinfo {year}
  {2008})}\BibitemShut {NoStop}%
\bibitem [{\citenamefont {Liechtenstein}\ \emph {et~al.}(1987)\citenamefont
  {Liechtenstein}, \citenamefont {Katsnelson}, \citenamefont {Antropov},\ and\
  \citenamefont {Gubanov}}]{liechtenstein1987local}%
  \BibitemOpen
  \bibfield  {author} {\bibinfo {author} {\bibfnamefont {A.~I.}\ \bibnamefont
  {Liechtenstein}}, \bibinfo {author} {\bibfnamefont {M.}~\bibnamefont
  {Katsnelson}}, \bibinfo {author} {\bibfnamefont {V.}~\bibnamefont
  {Antropov}}, \ and\ \bibinfo {author} {\bibfnamefont {V.}~\bibnamefont
  {Gubanov}},\ }\href@noop {} {\bibfield  {journal} {\bibinfo  {journal}
  {Journal of Magnetism and Magnetic Materials}\ }\textbf {\bibinfo {volume}
  {67}},\ \bibinfo {pages} {65} (\bibinfo {year} {1987})}\BibitemShut {NoStop}%
\bibitem [{\citenamefont {Hoffmann}\ \emph {et~al.}(2020)\citenamefont
  {Hoffmann}, \citenamefont {{Arthur Ernst}}, \citenamefont {Hergert},
  \citenamefont {Antonov}, \citenamefont {Adeagbo}, \citenamefont {Geilhufe},\
  and\ \citenamefont {{Ben Hamed}}}]{Hoffmann2020pssb}%
  \BibitemOpen
  \bibfield  {author} {\bibinfo {author} {\bibfnamefont {M.}~\bibnamefont
  {Hoffmann}}, \bibinfo {author} {\bibnamefont {{Arthur Ernst}}}, \bibinfo
  {author} {\bibfnamefont {W.}~\bibnamefont {Hergert}}, \bibinfo {author}
  {\bibfnamefont {V.~N.}\ \bibnamefont {Antonov}}, \bibinfo {author}
  {\bibfnamefont {W.~A.}\ \bibnamefont {Adeagbo}}, \bibinfo {author}
  {\bibfnamefont {R.~M.}\ \bibnamefont {Geilhufe}}, \ and\ \bibinfo {author}
  {\bibfnamefont {H.}~\bibnamefont {{Ben Hamed}}},\ }\href@noop {} {\bibfield
  {journal} {\bibinfo  {journal} {physica status solidi (b)}\ } (\bibinfo
  {year} {2020})}\BibitemShut {NoStop}%
\bibitem [{\citenamefont {Vitos}\ \emph {et~al.}(1994)\citenamefont {Vitos},
  \citenamefont {Koll{\'a}r},\ and\ \citenamefont {Skriver}}]{vitos1994full}%
  \BibitemOpen
  \bibfield  {author} {\bibinfo {author} {\bibfnamefont {L.}~\bibnamefont
  {Vitos}}, \bibinfo {author} {\bibfnamefont {J.}~\bibnamefont {Koll{\'a}r}}, \
  and\ \bibinfo {author} {\bibfnamefont {H.~L.}\ \bibnamefont {Skriver}},\
  }\href@noop {} {\bibfield  {journal} {\bibinfo  {journal} {Physical Review
  B}\ }\textbf {\bibinfo {volume} {49}},\ \bibinfo {pages} {16694} (\bibinfo
  {year} {1994})}\BibitemShut {NoStop}%
\end{thebibliography}%


\begin{thebibliography}{3}%
\makeatletter
\providecommand \@ifxundefined [1]{%
 \@ifx{#1\undefined}
}%
\providecommand \@ifnum [1]{%
 \ifnum #1\expandafter \@firstoftwo
 \else \expandafter \@secondoftwo
 \fi
}%
\providecommand \@ifx [1]{%
 \ifx #1\expandafter \@firstoftwo
 \else \expandafter \@secondoftwo
 \fi
}%
\providecommand \natexlab [1]{#1}%
\providecommand \enquote  [1]{``#1''}%
\providecommand \bibnamefont  [1]{#1}%
\providecommand \bibfnamefont [1]{#1}%
\providecommand \citenamefont [1]{#1}%
\providecommand \href@noop [0]{\@secondoftwo}%
\providecommand \href [0]{\begingroup \@sanitize@url \@href}%
\providecommand \@href[1]{\@@startlink{#1}\@@href}%
\providecommand \@@href[1]{\endgroup#1\@@endlink}%
\providecommand \@sanitize@url [0]{\catcode `\\12\catcode `\$12\catcode
  `\&12\catcode `\#12\catcode `\^12\catcode `\_12\catcode `\%12\relax}%
\providecommand \@@startlink[1]{}%
\providecommand \@@endlink[0]{}%
\providecommand \url  [0]{\begingroup\@sanitize@url \@url }%
\providecommand \@url [1]{\endgroup\@href {#1}{\urlprefix }}%
\providecommand \urlprefix  [0]{URL }%
\providecommand \Eprint [0]{\href }%
\providecommand \doibase [0]{http://dx.doi.org/}%
\providecommand \selectlanguage [0]{\@gobble}%
\providecommand \bibinfo  [0]{\@secondoftwo}%
\providecommand \bibfield  [0]{\@secondoftwo}%
\providecommand \translation [1]{[#1]}%
\providecommand \BibitemOpen [0]{}%
\providecommand \bibitemStop [0]{}%
\providecommand \bibitemNoStop [0]{.\EOS\space}%
\providecommand \EOS [0]{\spacefactor3000\relax}%
\providecommand \BibitemShut  [1]{\csname bibitem#1\endcsname}%
\let\auto@bib@innerbib\@empty
\bibitem [{\citenamefont {Hirahara}\ \emph {et~al.}(2017)\citenamefont
  {Hirahara}, \citenamefont {Eremeev}, \citenamefont {Shirasawa}, \citenamefont
  {Okuyama}, \citenamefont {Kubo}, \citenamefont {Nakanishi}, \citenamefont
  {Akiyama}, \citenamefont {Takayama}, \citenamefont {Hajiri}, \citenamefont
  {Ideta} \emph {et~al.}}]{hirahara2017large}%
  \BibitemOpen
  \bibfield  {author} {\bibinfo {author} {\bibfnamefont {T.}~\bibnamefont
  {Hirahara}}, \bibinfo {author} {\bibfnamefont {S.~V.}\ \bibnamefont
  {Eremeev}}, \bibinfo {author} {\bibfnamefont {T.}~\bibnamefont {Shirasawa}},
  \bibinfo {author} {\bibfnamefont {Y.}~\bibnamefont {Okuyama}}, \bibinfo
  {author} {\bibfnamefont {T.}~\bibnamefont {Kubo}}, \bibinfo {author}
  {\bibfnamefont {R.}~\bibnamefont {Nakanishi}}, \bibinfo {author}
  {\bibfnamefont {R.}~\bibnamefont {Akiyama}}, \bibinfo {author} {\bibfnamefont
  {A.}~\bibnamefont {Takayama}}, \bibinfo {author} {\bibfnamefont
  {T.}~\bibnamefont {Hajiri}}, \bibinfo {author} {\bibfnamefont {S.-i.}\
  \bibnamefont {Ideta}},  \emph {et~al.},\ }\href@noop {} {\bibfield  {journal}
  {\bibinfo  {journal} {Nano letters}\ }\textbf {\bibinfo {volume} {17}},\
  \bibinfo {pages} {3493} (\bibinfo {year} {2017})}\BibitemShut {NoStop}%
\bibitem [{\citenamefont {Xie}\ \emph {et~al.}(2003)\citenamefont {Xie},
  \citenamefont {Liu},\ and\ \citenamefont {Pettifor}}]{xie2003half}%
  \BibitemOpen
  \bibfield  {author} {\bibinfo {author} {\bibfnamefont {W.-H.}\ \bibnamefont
  {Xie}}, \bibinfo {author} {\bibfnamefont {B.-G.}\ \bibnamefont {Liu}}, \ and\
  \bibinfo {author} {\bibfnamefont {D.}~\bibnamefont {Pettifor}},\ }\href@noop
  {} {\bibfield  {journal} {\bibinfo  {journal} {Physical Review B}\ }\textbf
  {\bibinfo {volume} {68}},\ \bibinfo {pages} {134407} (\bibinfo {year}
  {2003})}\BibitemShut {NoStop}%
\bibitem [{\citenamefont {Eremeev}\ \emph {et~al.}(2017)\citenamefont
  {Eremeev}, \citenamefont {Otrokov},\ and\ \citenamefont
  {Chulkov}}]{eremeev2017competing}%
  \BibitemOpen
  \bibfield  {author} {\bibinfo {author} {\bibfnamefont {S.}~\bibnamefont
  {Eremeev}}, \bibinfo {author} {\bibfnamefont {M.}~\bibnamefont {Otrokov}}, \
  and\ \bibinfo {author} {\bibfnamefont {E.}~\bibnamefont {Chulkov}},\
  }\href@noop {} {\bibfield  {journal} {\bibinfo  {journal} {Journal of Alloys
  and Compounds}\ }\textbf {\bibinfo {volume} {709}},\ \bibinfo {pages} {172}
  (\bibinfo {year} {2017})}\BibitemShut {NoStop}%
\end{thebibliography}%

\end{document}


\title{Supplementary Information for: Domain wall induced spin-polarized flat bands in antiferromagnetic topological insulators}

\author{E. K. Petrov}
\email[]{evg.konst.petrov@gmail.com}
\affiliation{Tomsk State University, 634050 Tomsk, Russia}
\affiliation{St. Petersburg State University, 199034 St. Petersburg,  Russia}

\author{V. N. Men'shov}
\affiliation{NRC Kurchatov Institute, 123182 Moscow, Russia}
\affiliation{Tomsk State University, 634050 Tomsk, Russia}
\affiliation{St. Petersburg State University, 199034 St. Petersburg,  Russia}

\author{I. P. Rusinov}
\affiliation{Tomsk State University, 634050 Tomsk, Russia}
\affiliation{St. Petersburg State University, 199034 St. Petersburg,  Russia}

\author{M. Hoffmann}
\affiliation{Institute for Theoretical Physics, Johannes Kepler University, 4040 Linz, Austria}

\author{A. Ernst}
\affiliation{Institute for Theoretical Physics, Johannes Kepler University, 4040 Linz, Austria}
\affiliation{Max Planck Institute of Microstructure Physics, 06120 Halle, Germany}

\author{M. M. Otrokov}
\affiliation{Centro de F\'{\i}sica de Materiales (CFM-MPC), Centro Mixto CSIC-UPV/EHU, 20018 Donostia-San Sebasti\'{a}n, Basque Country, Spain}
\affiliation{IKERBASQUE, Basque Foundation for Science, 48011 Bilbao, Spain}

\author{V. K. Dugaev}
\affiliation{Department of Physics and Medical Engineering, Rzesz\'ow University of Technology, 35-959 Rzesz\'ow, Poland}

\author{T. V. Menshchikova}
\affiliation{Tomsk State University, 634050 Tomsk, Russia}

\author{E. V. Chulkov}
\affiliation{Departamento de F\'{\i}sica de Materiales, Facultad de Ciencias Qu\'{\i}micas, Universidad del Pa\'{\i}s Vasco, Apdo. 1072, 20080 San Sebasti\'{a}n/Donostia, Spain}
\affiliation{Donostia International Physics Center (DIPC), 20018 San Sebasti\'{a}n/Donostia, Spain}
\affiliation{St. Petersburg State University, 199034 St. Petersburg,  Russia}
\affiliation{Centro de F\'{\i}sica de Materiales (CFM-MPC), Centro Mixto CSIC-UPV/EHU, 20018 Donostia-San Sebasti\'{a}n, Basque Country, Spain}
\affiliation{Tomsk State University, 634050 Tomsk, Russia}

\newcommand*{\vv}[1]{\textbf{#1}}
\newcommand*{\BT}{Bi$_2$Te$_3$}
\newcommand*{\BS}{Bi$_2$Se$_3$}
\newcommand*{\ST}{Sb$_2$Te$_3$}
\newcommand*{\STS}{Sb$_2$Te$_2$Se}
\newcommand*{\BTS}{Bi$_2$Te$_2$Se}
\newcommand*{\MBS}{MnBi$_2$Se$_4$}
\newcommand*{\MBT}{MnBi$_2$Te$_4$}
\newcommand*{\MBTS}{MnBi$_2$Te$_2$Se$_2$}
\newcommand*{\VBS}{VBi$_2$Se$_4$}
\newcommand*{\VBTS}{VBi$_2$Te$_2$Se$_2$}
\newcommand*{\VBT}{VBi$_2$Te$_4$}
\newcommand*{\VST}{VSb$_2$Te$_4$}
\newcommand*{\VSTS}{VSb$_2$Te$_2$Se$_2$}
\date{\today}

\maketitle

\section*{Supplementary Note 1: Equllibrium crystal structure and formation energies}
\begin{figure}[h!]
	\includegraphics[width=8cm]{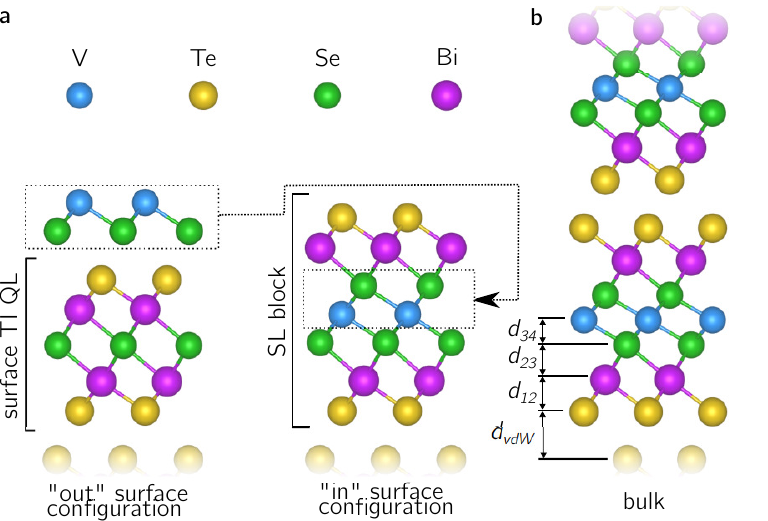}%
	\caption{\label{fig:formation} (\textbf{a}) \VBTS{} SL formation scheme. (\textbf{b}) Bulk interplane distances notation.}
\end{figure}

According to the last reports \cite{hirahara2017large}, MnSe bilayer tends to diffuse from \BS{} surface to the center of topmost TI QL, forming an \MBS{} SL.
In order to show, that similar process is possible for monochalcogenides VSe, VTe and MnSe deposed on the surface of a typical TI (\BS{}, \BTS{}, \BT{}, \STS{} or \ST{}) surface, we focus on total energy comparison of ``out'' (deposed bilayer is located on the TI QL surface) and ``in'' (deposed bilayer is located in TI QL center) configurations for VSe/\BS{}, VSe/\BTS{}, VTe/\BT{}, VSe/\STS{}, VTe/\ST{} and MnSe/\BTS{} pairs without considering the diffusion mechanism of deposited atoms inside the QL (Suppl. Fig. \ref{fig:formation}).
In ``out'' configuration the deposed bilayer can be oriented in two ways: forming V or Se (Te) termination, respectively.
We found V termination to be preferable for VSe/\BS{} and VSe/\BTS{} cases, while VTe/\BT{} prefers Te one.
Comparison of the total energy of ``out'' and ``in'' configurations clearly indicates that VSe (VTe) diffusion to the topmost QL center is more favorable (Table \ref{table:formation}).

Note, one of the necessary conditions for introducing a magnetic monochalcogenide bilayer to the center of topmost TI structural block is small in-plane lattice mismatch of the bilayer and substrate. 
It was shown that the V-based monochalcogenides can exhibit hexagonal NiAs-type (NA) and wurtzite-type (WZ) phases but the last one was not yet obtained experimentally. 
However, using ab-initio approach, their equllibrium lattice parameters were calculated \cite{xie2003half}. 
Similarly to the MnSe, resulting WZ phase lattice parameters of V-based monochalcogenides match with substrate much better than NA ones and guarantee the possibility of well structurized SL formation.

Since a magnetic SL can be formed on a TI surface, thick magnetic films can be also constructed.
For example, VSe deposition on \BS{} surface will create one \VBS{} SL. After that, one \BS{} QL may be deposed
over \VBS{} SL, and this process may be repeated. By doing so, it is possible to grow a magnetic film thick
enough to bulk electronic properties staring to emerge. Hence, bulk crystal and electronic structure
investigation is vital for a complete analysis of compounds under study. 

As it was mentioned in the main text,  in the case of \VBS{} total energy difference between rhombohedral and monoclinic structure is about 100 times less than in all other considered cases. Since at this stage all compounds were assumed to be FM, such small energy profit may indicate that monoclinic magnetic ground state may be  AFM. We considered 4 different types of AFM order (AFM-1, AFM-2, AFM-3, AFM-4, see Ref. \cite{eremeev2017competing}) of monoclinic \VBS{}, and found its magnetic ground state to be FM. Considered AFM orders were found to exhibit total energies higher than FM one by 0.307~meV/f.u., 5.09~meV/f.u., 5.09~meV/f.u. and 5.244~meV/f.u. for AFM-1, AFM-2, AFM-3, AFM-4 orders, respectively. 

\begin{table}[h!]
	\caption{
		\label{table:formation} Total energy difference between ``out'' ($E_{out}$) and ``in'' ($E_{in}$) surface configurations.  Positive value means that ``in'' configuration is favorable. First element in the first column indicated preferable termination in ``out'' configuration (e.g., TeV/\BT{} prefers Te termination and VSe/\BTS{} prefers Se termination as it shown on Suppl. Fig.~\ref{fig:formation}a). 	}
	\begin{ruledtabular}
		\begin{tabular}{cc}
			System   & $E_{out} - E_{in} (eV/f.u.)$ \\ \hline
			MnSe/\BTS &       0.470        \\ \hline
			VSe/\BS  &       1.063        \\ \hline
			VSe/\BTS  &       0.983        \\ \hline
			TeV/\BT  &       0.861       \\ \hline
			TeV/\ST  &      0.95   \\ \hline
			SeV/\STS  &     1.021
		\end{tabular}
	\end{ruledtabular}
\end{table}
\begin{table}[h!]
	\caption{\label{table:bulk-cryst} Total energy difference between monoclinic ($E_{min}$) and rhombohedral ($E_{rhomb}$) bulk phases. The energy units are meV/f.u. Positive value means that rhombohedral bulk phase is favorable.}
	\begin{ruledtabular}
		\begin{tabular}{cc}
			System & $E_{mon} - E_{rhomb}$ (meV/f.u.) \\ \hline
			\MBTS  &       $+225.5$   \\ \hline
			\VBS  &        $+0.9$    \\ \hline
			\VBTS  &       $+318.3$   \\ \hline
			\VBT  &       $+180.6$   \\ \hline
			\VSTS  &       $+239.8$   \\ \hline
			\VST  &       $+116.8$  
		\end{tabular}
	\end{ruledtabular}
\end{table}%
\begin{table}[h!]
	\caption{\label{table:bulk-cryst-param} Lattice parameters and interlayer distances of rhombohedral bulk plases. All units are angstroms.}
	\begin{ruledtabular}
		\begin{tabular}{cccccc}
			System & $a_{hex}$ & $d_{12}$  & $d_{23}$  & $d_{34}$  & $d_{vdW}$ \\ \hline
			\MBTS  & $4.1904$  & $1.82863$ & $1.95850$ & $1.40135$ & $2.70315$ \\ \hline
			\VBS  & $4.0783$  & $1.60575$ & $1.99966$ & $1.42537$ & $2.49110$ \\ \hline
			\VBTS  & $4.1906$  & $1.82349$ & $1.94649$ & $1.38162$ & $2.75384$ \\ \hline
			\VBT  & $4.3380$  & $1.75433$ & $2.11106$ & $1.53298$ & $2.59393$ \\ \hline
			\VSTS  & $4.1124$  & $1.76196$ & $1.86362$ & $1.41252$ & $2.88337$ \\ \hline
			\VST  & $4.2605$  & $1.69003$ & $2.02189$ & $1.58758$ & $2.69557$
		\end{tabular}
	\end{ruledtabular}
\end{table}%

\section*{Supplementary Note 2: Band structures}
\begin{figure*}[]
	\centering
	\includegraphics{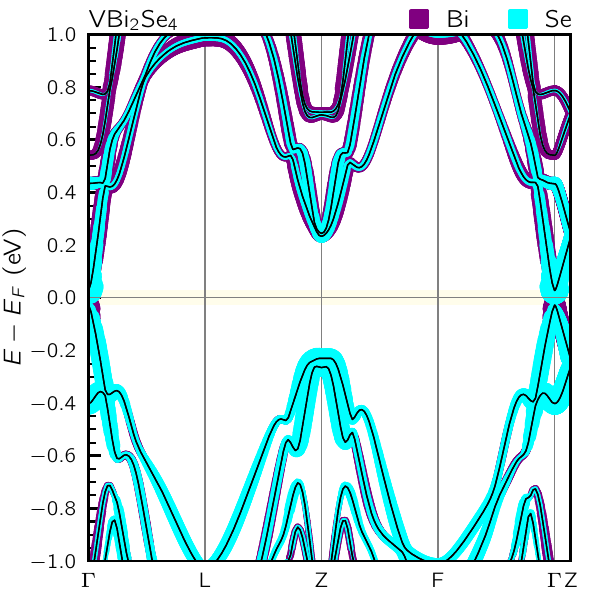}
	\includegraphics{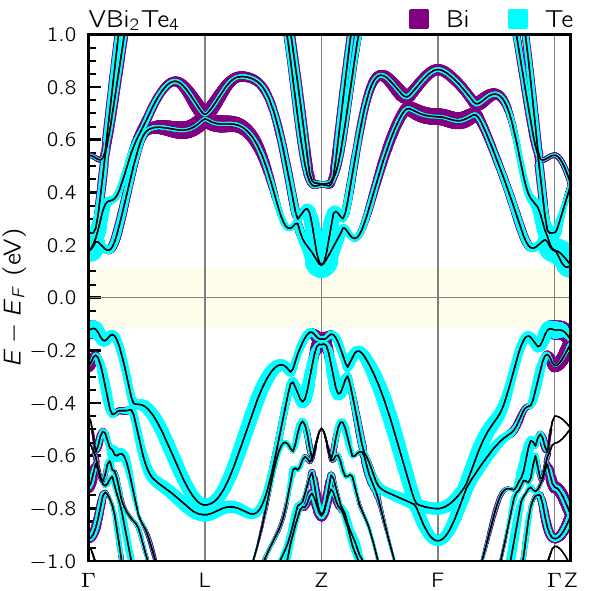}
	\includegraphics{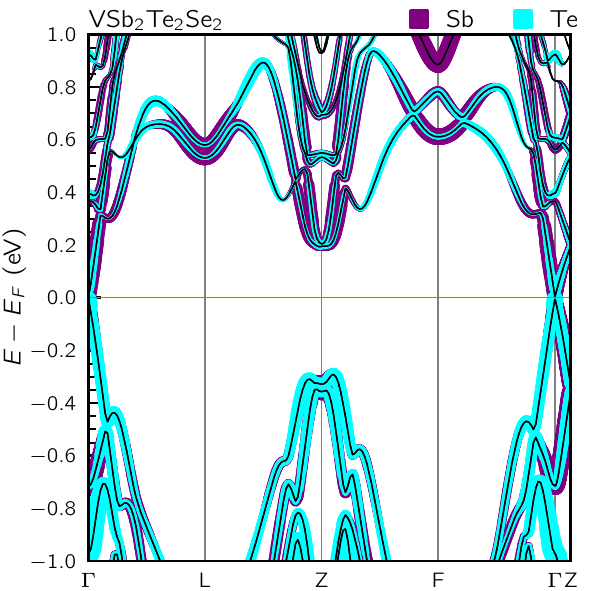}
	\includegraphics{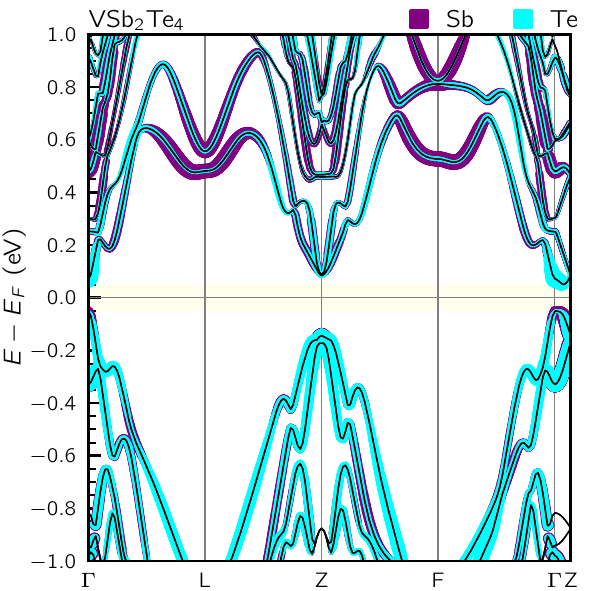}
	\includegraphics{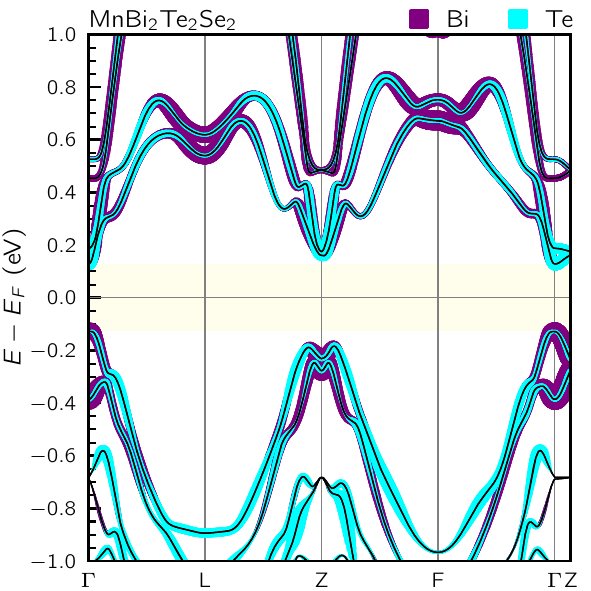}
	\caption{\label{fig:bnd_bulk} \VBS{}, \VBT{}, \VSTS{}, \VST{} and \MBTS{} bulk band structures near the Fermi level. Notations are the same as in Fig.~2c in the main text.}
	
\end{figure*}

\begin{figure*}[]
	\centering
	\includegraphics{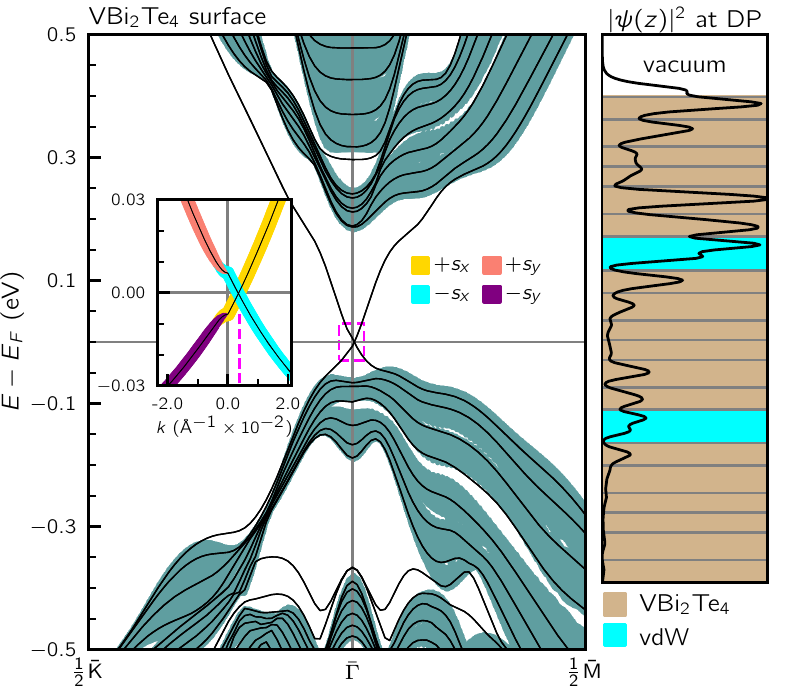}
	\includegraphics{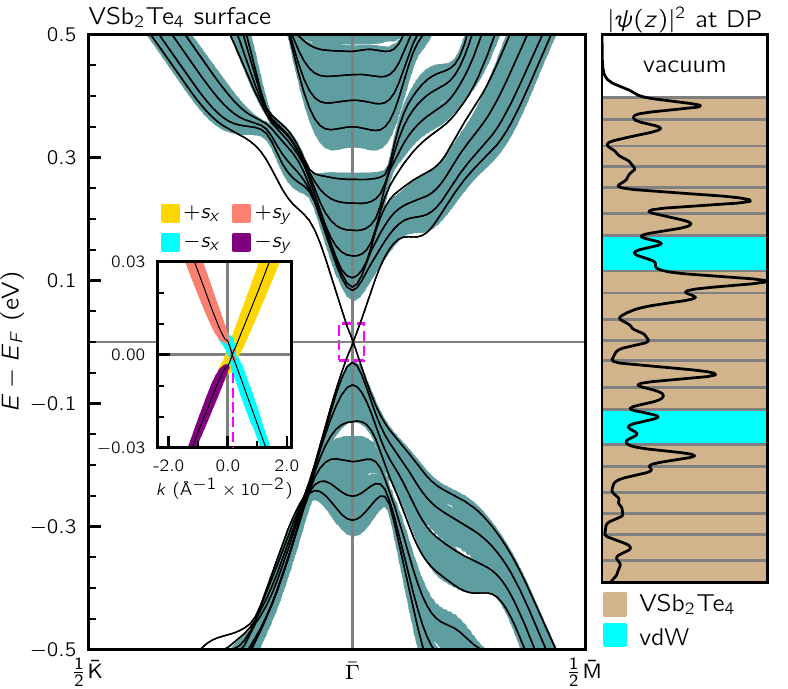}
	\includegraphics{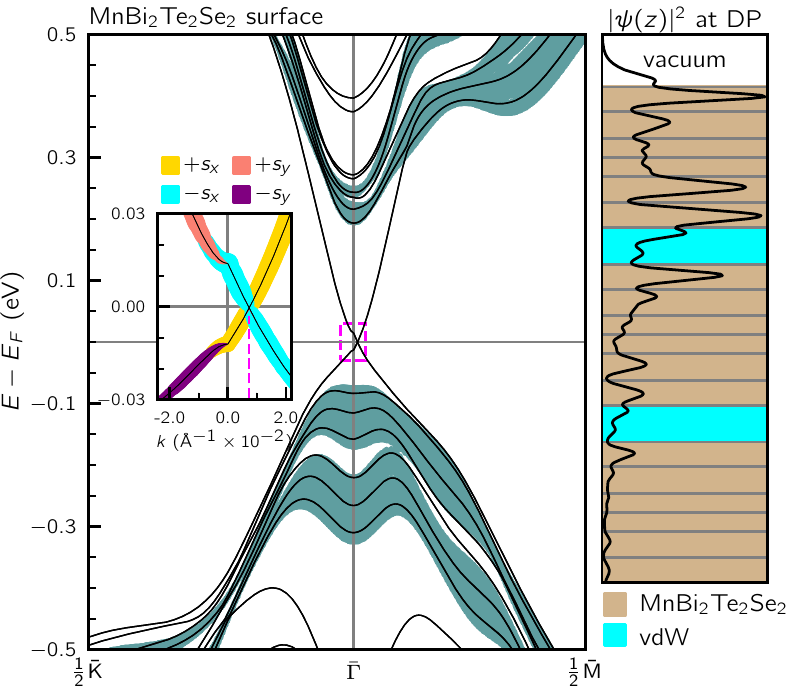}
	\caption{\label{fig:bnd_surf} \VBT{}, \VST{} and \MBTS{} surface band structures. Notations are the same as in Fig.~3a in the main text.}
\end{figure*}

Here we present the band structures, which were not shown in the main text (see Suppl. Fig.~\ref{fig:bnd_bulk} for bulk and Suppl. Fig.~\ref{fig:bnd_surf} for surface band structures, respectively). We do not show surface band structures of compounds which exhibit $\mathbb{Z}_2=0$ (see Table in the main text).

\bibliography{inplane_AFMTI_bib}